\documentclass[twocolumn,tighten]{aastex61}
\usepackage{amsmath}
\usepackage{graphicx}
\usepackage{color}
\usepackage{amsmath} % or simply amstext

\newcommand{\cmc}{{\tt{CMC}}}
\newcommand{\sse}{{\tt SSE}}
\newcommand{\bse}{{\tt BSE}}
\newcommand{\msun}{{\rm{M_\odot}}}

\usepackage{hyperref}
\usepackage{longtable}

\usepackage{filecontents}

\begin{document}
\bibliographystyle{apj}

\title{Low-mass x-ray binaries ejected from globular clusters}

%% Authors with the same affiliation can be grouped in a single
%% \author and \affil call.
\author[0000-0002-4086-3180]{Kyle Kremer}
\affil{\centering Center for Interdisciplinary Exploration and Research in Astrophysics (CIERA), Evanston, IL 60208, USA}
\affil{\centering Department of Physics \& Astronomy, Northwestern University, Evanston, IL 60208, USA}
\email{kremer@u.northwestern.edu}
\author[0000-0002-3680-2684]{Sourav Chatterjee}
\affil{\centering Center for Interdisciplinary Exploration and Research in Astrophysics (CIERA), Evanston, IL 60208, USA}
\affil{\centering Department of Physics \& Astronomy, Northwestern University, Evanston, IL 60208, USA}
\affil{\centering Tata Institute of Fundamental Research, Homi Bhabha Road, Mumbai 400005, India}
\email{chatterjee.sourav2010@gmail.com}
\author{Carl L. Rodriguez}
\affil{\centering MIT-Kavli Institute for Astrophysics and Space Research, Cambridge, MA 02139, USA}
\author[0000-0002-7132-418X]{Frederic A. Rasio}
\affil{\centering Center for Interdisciplinary Exploration and Research in Astrophysics (CIERA), Evanston, IL 60208, USA}
\affil{\centering Department of Physics \& Astronomy, Northwestern University, Evanston, IL 60208, USA}

\begin{abstract}
We explore the population of mass-transferring binaries ejected from globular clusters (GCs) with both black hole (BH) and neutron star (NS) accretors. We use a set of 137 fully evolved globular cluster models  which span a large range in cluster properties and, overall, match very well the properties of old GCs observed in the Milky Way. We identify all binaries ejected from our set of models that eventually undergo mass-transfer. These binaries are ejected from their host clusters over a wide range of ejection times and include white dwarf, giant, and main sequence donors. We calculate the orbits of these ejected systems in the Galactic potential to determine their present-day positions in the Galaxy and compare to the distribution of observed low-mass X-ray binaries (XRBs) in the Milky Way. We estimate $\sim 300$ mass-transferring NS binaries and $\sim 180$ mass-transferring BH binaries may currently be present in the Milky Way that originated from within GCs. Of these, we estimate, based on mass-transfer rates and duty cycles at the present time,  at most a few would be observable as BH--XRBs and NS--XRBs at the present day. Based on our results, XRBs that originated from GCs are unlikely to contribute significantly to the total population of low-mass XRBs in the Galactic field.
\end{abstract}
\keywords{X-rays: binaries–-globular clusters: general--stars: black holes-–stars: neutron--stars: kinematics and dynamics--methods: numerical}

\section{Introduction} \label{sec:intro}
Low-mass X-ray binaries (LMXBs) are binary systems in which a neutron star (NS) or black hole (BH) accretes matter from a low-mass ($\lesssim 1.5 M_{\odot}$) companion star. Several hundreds of LMXBs have been observed in our galaxy \citep[e.g.,][]{Liu2007, CorralSantana2016, Jonker2004}. 

Many of these LMXBs are found within globular clusters (GCs). Prior to 2007, all XRBs discovered in GCs had NS accretors. Recently, mass-transferring BH binary candidates have been discovered in both extragalactic \citep[e.g.,][]{Maccarone2007,Irwin2010} and Galactic \citep[e.g.,][]{Strader2012,Chomiuk2013,Miller-Jones2014} GCs. 

Over the course of the evolution of a GC, stars will be ejected as the result of natal kicks associated with supernovae (SNe), tidal stripping from the host galaxy, and dynamical encounters. If binary systems containing a BH/NS and a luminous companion are ejected from GCs through these channels, GCs may contribute to the population of XRBs found in the halo of their host galaxy \citep[e.g.,][]{Giesler2017}.

The formation mechanism for LMXBs in GCs can be dramatically different relative to that in the galactic field; inside GCs, most LMXBs are dynamically assembled \citep[e.g.,][]{Ivanova2010,Naoz2016,Kremer2017}, in contrast, the field population is created from primordial binaries.
%As a result, \citet{Giesler2017} suggested, at least in the case of BH binaries, the mass-transferring systems that originated inside GCs may be distinguishable from those formed in the Galactic field \authorcomment1{Add something like "by comparing e, separations, etc. basically some properties"}. 
%Previous studies \citep[e.g.,][]{Giesler2017} have noted that, in the case of BH binaries, the mass-transferring systems which originated in GCs may be distinguishable from those systems which originated within the Galactic field as the result of the role dynamical processes play in the formation and evolution of these systems.

Dynamical interactions are likely less important for NS--XRBs relative to BH--XRBs. It is expected that NSs receive larger kicks at birth, relative to BHs, and as a result, a higher fraction of NSs are ejected from their host cluster before dynamical interactions can take place. However, several previous studies \citep[e.g.,][]{Clark1975, Bailyn1995, Heinke2010} have noted that there is an overabundance of NS--XRBs observed in GCs relative to the Galactic field, suggesting that dynamical processes may nonetheless play an important role in the formation and evolution of those NS binaries that were not ejected from the GCs via natal kicks.
%NSs are not significantly more massive than the average cluster object, and therefore will not rapidly mass-segregate upon formation like the BHs. Therefore, NSs are expected to experience fewer dynamical encounters, on average, than BHs, which means the evolution of NS binaries is more sensitive to the highly uncertain parameters of binary star evolution, for example common envelope and natal kick prescriptions.

In this analysis, we explore the properties of all mass-transferring binaries with BH and NS accretors {\em ejected} from GCs. We use a set of 137 fully-evolved GC models, produced using our \texttt{CMC} cluster dynamics code. In Section \ref{sec:method}, we discuss the methods used to calculate the evolution of GCs as well as the evolution of binary systems after they are ejected from their host cluster. We also discuss the various ways these systems can be ejected from their host cluster. In Section \ref{sec:ejected_systems} we discuss the total numbers of BH and NS binaries ejected from our GC models which will eventually become mass-transferring systems, and discuss the various features of these systems. In Section \ref{sec:gx-ev}, we introduce a method to calculate the orbital evolution of all ejected systems within the Galactic potential in order to determine the present-day $z$-distribution of these systems within the Galactic halo, and compare to the $z$-distribution of observed XRBs in the Milky Way (MW). In Section \ref{sec:luminosity}, we discuss our method for calculating the X-ray luminosities and duty cycles of these mass-transferring systems and determine which of these systems may actually be observable as X-ray sources at the present-day. We conclude in Section \ref{sec:conclusion}.

\section{Method}
\label{sec:method}

\subsection{Modeling globular cluster evolution}
\label{sec:model_properties}
We use our H\'{e}non-style Monte Carlo cluster dynamics code, \cmc, to model massive star clusters 
\citep[e.g.,][]{Joshi2000,Joshi2001,Fregeau2003, Fregeau2007, Chatterjee2010,Chatterjee2013,Umbreit2012,Pattabiraman2013,Morscher2013,Morscher2015,Rodriguez2016b}. 

\cmc\ incorporates all relevant physical processes for studying the formation and evolution of binary systems containing BHs and NSs including two-body relaxation, binary-mediated gravitational scattering encounters \citep[modeled explicitly using the {\tt Fewbody} small-$N$ integrator;][]{Fregeau2004}, and single and binary stellar evolution \citep[implemented using the \sse\ and \bse\ 
software packages; ][]{Hurley2000,Hurley2002,Chatterjee2010,Kiel2009}. %and modified to incorporate our modern understanding of the BH mass function, stellar winds, and natal kicks due to supernovae (SN)  \citep[e.g.,][]{Vink2001,Fryer2001,Belczynski2002}).

We use $137$ independent GC models, which are listed in Table \ref{table:params} in the Appendix. Several different initial cluster properties are varied, including the initial number of cluster objects ($N$), the initial galactocentric distance ($r_{\rm{G}}$), the King concentration parameter ($w_o$), the overall primordial binary fraction ($f_b$), the initial metallicity ($Z$), and the initial virial radius ($r_v$). The initial values of these parameters for each model are shown in columns 2--7, respectively, of Table \ref{table:params}.

All core-collapsed NSs receive birth kicks drawn from a Maxwellian distribution with $\sigma=\sigma_{\rm{NS}}=265\, \rm{km\,s}^{-1}$ \citep{Hobbs2005}. In the case of NSs formed through accretion-induced collapse of  a white dwarf, birth kicks are drawn from a Maxwellian distribution with $\sigma_{\rm{AIC}} = 20 \, \rm{km\,s}^{-1}$

One major uncertainty is related to the natal kicks BHs receive which lacks strong observational or theoretical constraints. Hence, we remain agnostic to the ``right'' prescription for BH natal kicks and instead use four different prescriptions to calculate magnitudes for natal kicks imparted to BHs. In the first prescription, we assume BHs are formed with significant fallback and calculate the natal kicks by sampling from the same kick distribution as the NSs, but reduced in magnitude according to the fractional mass of the fallback material 
\citep[e.g.,][]{Fryer2001,Belczynski2002,Morscher2015}. Models using this prescription are marked ``y" in column 9 of Table \ref{table:params} (marked $\rm{FB}$) denoting that fallback is turned on. In the other three variations, we neglect fallback and simply use $\sigma_{\rm{BH}}=\sigma_{\rm{NS}}$, $\sigma_{\rm{BH}}=0.1\,\sigma_{\rm{NS}}$, and $\sigma_{\rm{BH}}=0.01 \, \sigma_{\rm{NS}}$. Models using these prescriptions are marked ``n" in column 9 of Table \ref{table:params} and the ratio of $\sigma_{\rm{BH}}$ to $\sigma_{\rm{NS}}$ for each model is given in column 8 of Table \ref{table:params}.

The initial stellar masses (primary masses for primordial binaries) are sampled from the initial mass function (IMF) given in \citet{Kroupa2001}. The IMF range is chosen as either $0.1$--$100\,\msun$ or 0.08--150 $\msun$ (listed in column 13 of Table \ref{table:params}). 
%and $\alpha_1=2.3$ is used for all models. 
An appropriate number of stars are then randomly chosen based on the adopted $f_b$ and $N$ for each model. For binary systems, secondary masses are assigned based on a flat distribution in mass ratios ($q\equiv m_s/m_p$, where $m_s$ and $m_p$ denote the secondary and primary masses, respectively). The initial orbital periods ($P$) of binaries are drawn from a distribution of the form $dn/d\log P\propto P^\alpha$ and the initial eccentricities are thermal.

In some models, we specifically vary the initial binary fraction, $f_{b,\rm{high}}$, for high-mass stars ($>15\,\msun$) independent of the overall binary fraction. In these models, we also vary the range in $q$ and the initial period distribution for the high-mass stars motivated by the 
observational constraints from \citet[][]{Sana2012}. Columns 10--12 of Table \ref{table:params} show the initial binary fraction, mass ratio range, and period distribution for high mass stars for each model.

To calculate the common envelope (CE) evolution for binary systems, all our models use the $\alpha \lambda$-CE prescription detailed in \cite{Hurley2002} with $\lambda$ calculated using a fitting formula similar to that described in \citet{Claeys2014}.%with the exception of two models (models 2 and 3 in Table \ref{table:params}, marked with a $*$). For model 2, we use $\alpha \lambda = 0.01$, representing high CE binding energy and for model 3, we use $\alpha \lambda = 10$, representing low CE binding energy.

To treat mass-loss due to stellar winds, we use two prescriptions: In the first prescription, we use the method described in \citet{Hurley2000}, as implemented in \texttt{SSE} and \texttt{BSE}. We label this wind implementation as the ``strong wind'' prescription, and mark such models with an ``S'' in column 14 of Table \ref{table:params}. In the second prescription, we used the method detailed in \citet{Belczynski2010b} which reflects the recent observations of high-mass stars that suggest winds may not be as strong as suggested by earlier studies \citep[e.g.,][]{Vink2001,Vink2008,Belczynski2010a,Belczynski2010b,Dominik2012,Spera2015}. We label this wind implementation as the ``weak wind'' prescription, and mark such models with a ``W'' in column 14 of Table \ref{table:params}.

For each of these 137 models, we identify all compact object binaries ejected from each cluster with luminous companions and determine whether each binary will ever undergo mass-transfer by evolving it forward as an isolated system. We follow the binary evolution of each ejected binary after ejection using the same version of \texttt{BSE} used in \texttt{CMC}. Additionally, we calculate the orbit of each ejected binary in the Galactic potential after ejection, discussed further in Section \ref{sec:gx-ev}.

\subsection{Ejection Mechanisms}
\label{sec:Ejection_mechanisms}
Over the course of a GC's evolution, cluster objects can be ejected from the cluster through a variety of mechanisms. Of particular relevance to the BH and NS binaries considered here are ejection through dynamical encounters with other objects in the cluster and ejection from kicks imparted to the compact object during collapse of the compact object's progenitor.

\subsubsection{Non-dynamical ejections}
It is expected that upon formation, all NSs and some BHs receive kicks assumed to be generated by the asymmetric ejection of mass during the collapse of the compact object's progenitor.

NSs are expected to form through two distinct channels: the standard core-collapse (CC) SN channel resulting from the evolution of a high-mass star and the accretion-induced collapse (AIC) channel which results when a white dwarf (WD) reaches the Chandrasekhar limit through mass-transfer from a binary companion and implodes via AIC to form a NS \citep[e.g.,][]{Nomoto1979,Taam1986,Canal1990,Tauris2013,Ruiter2018}.

The formation of a NS through AIC is just one possible scenario that may result when an accreting WD reaches the Chandrasekhar limit. Such an event may also result in a Type Ia SN, which would leave behind no remnant.
Here, we use the treatment implemented in \texttt{BSE} to determine the outcome of accretion onto a WD. This treatment is described in detail in Section 2.6 of \citet{Hurley2002}.

Measurements of pulsar proper motions \citep[e.g.,][]{Lyne1994,Hansen1997,Hobbs2005} and studies of NS--XRBs \citep[e.g.,][]{Johnston1992,Fryer1997,Pfahl2002} suggest that, upon formation, NSs acquire kicks which may be either low or high velocity, depending upon the formation channel associated with the NS. High velocity kicks (often referred to as natal kicks) are generally associated with standard CC SN scenario while low velocity kicks are expected to result from the AIC formation scenario.
%Here, NS natal kicks are drawn from a Maxwellian distribution with $\sigma = 265 \, \rm{km\,s}^{−1}$. . In the case of AIC kicks, we use $\sigma = 20 \, \rm{km\,s}^{-1}$.

%For the case of BH formation, we use the prescription of \citet{Belczynski2002}, which is based on the calculations in \citet{Fryer2001}. In this prescription, BHs can form either through direct collapse (i.e., with no SN explosion) or through partial fallback of material that was initially expelled in a SN explosion, depending on the mass of the stellar core just before BH formation. For BH kicks, we follow the prescription of \citet{Belczynski2002} to reduce the BH kick magnitude (initially drawn from the NS kick distribution) according to the amount of material that falls back onto the final BH after the SN. 

%\authorcomment1{The discussion about kicks is repeated. it is already discussed as part of ICs. I suggest not making ejection mechanisms as prominent. you actually don't have much to say here. just briefly mention that some are ejected during birth, and some are ejected after many dynamical encounters when the recoil is sufficient to eject it. }

\begin{figure*}
\begin{center}
\includegraphics[width=1.0\textwidth]{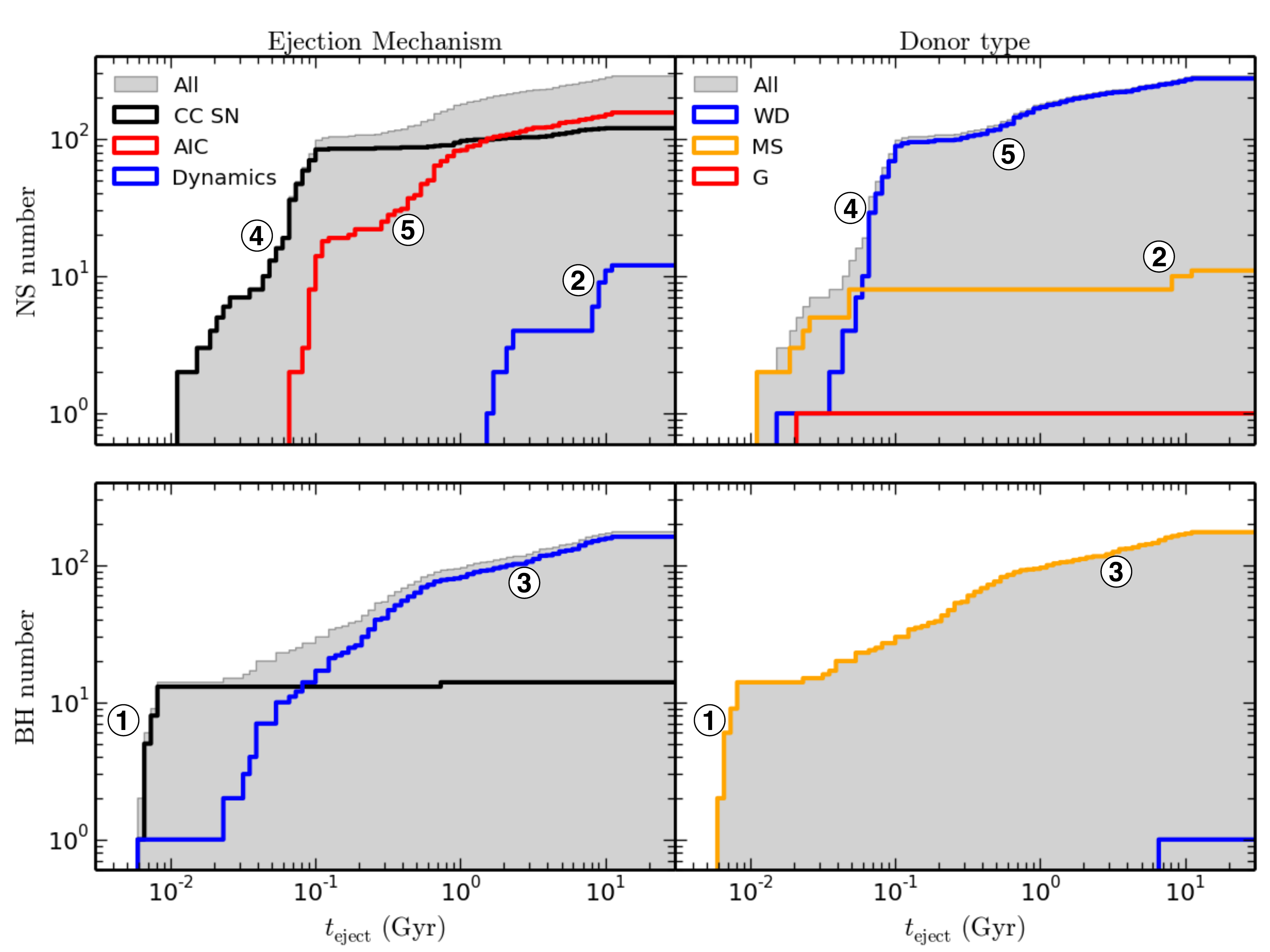}
\caption{\label{fig:ejection} Cumulative distribution of time of ejection for all ejected NS--MTBs (top panels) and BH--MTBs (bottom panels). The background gray histograms show the total number of systems. The different colors in the left-hand panels show the contributions to the total of the various ejection mechanisms. Blue indicates systems ejected by dynamical interactions, black marks systems ejected by CC SN, and red marks systems ejected by AIC. The different colors in the right-hand panels show the contributions to the total of the various donor types. Here, blue indicates systems with a WD donor, orange indicates systems with a MS donor, and red indicates systems with a giant donor. The circled numbers, 1--5, denote specific systems whose evolution is illustrated in detail in Figure \ref{fig:cartoon}.}
\end{center}
\end{figure*}

\subsubsection{Dynamical ejections}
For BH and NS binaries which remain in their host cluster after birth, strong or resonant dynamical encounters provide an additional mechanism to eject these binaries from their clusters. BHs, being more massive than the average cluster object, will rapidly mass-segregate upon formation \citep[e.g.,][]{Gurkan2004}. In the high-density regions near the cluster center, BH binaries are formed through three-body processes or binary--binary exchange encounters. Dynamical interactions between a compact binary and its surrounding components make the binary's orbit more compact \citep{Heggie1975, Goodman1993} and may provide significant kick velocities to the binary upon the conclusion of the interaction. Ejection from the host cluster occurs when their recoil velocity becomes greater than the escape velocity of the cluster.

The formation and ejection of BH binaries within the cores of GCs has been shown to be a key formation channel of BH--BH binaries that are gravitational wave sources for detectors such as LIGO \citep{Banerjee2010,Ziosi2014,Rodriguez2015,Rodriguez2016a,Chatterjee2017a,Chatterjee2017b}. BHs will also dynamically mix with non-BHs, potentially forming BH--non-BH binaries that may eventually mass-transfer and be observed as X-ray sources \citep[e.g.,][]{Kremer2017} within their host cluster. In a manner similar to the BH--BH binaries, these BH--non-BH binaries can also be ejected from their host clusters as the result of dynamical encounters.

The importance of dynamical interactions for NS binaries isn't as clear, as NSs do not mass-segregate as rapidly as the more massive BHs. However, the observed over abundance of NS--XRBs in GCs relative to the number of these systems  in the Galactic field suggests that dynamical interactions may play a crucial role in the formation of these systems within GCs, especially for NSs receiving low natal kicks and in binaries \citep[e.g.,][]{Clark1975}.

%\subsection{Modeling binary evolution after ejection}

%To be consistent with the binary evolution treatment implemented within \cmc, we use \bse to calculate the evolution of binaries after they are ejected from their host cluster. Upon ejection, the component masses, semi-major axis, eccentricity, and age (time of ejection) are given for each binary. Additionally, each GC model (and therefore each ejected binary) has a specified metallicity. We use these input parameters to evolve each binary forward in time from the time of ejection using \bse.

%Additionally, we calculate the orbit of each ejected binary in a Galactic potential after ejection. This is discussed further in Section \ref{sec:gx-ev}.

%Depending upon the stellar type  of the companion star at the time of ejection as well as the orbital parameters of each ejected system, these ejected binaries may or may not mass-transfer within a Hubble time. We discuss the total number of mass-transferring binaries in the following section.

\section{Ejected mass-transferring binaries}
\label{sec:ejected_systems}

\begin{figure*}
\begin{center}
\includegraphics[width=1.2\textwidth, angle=270]{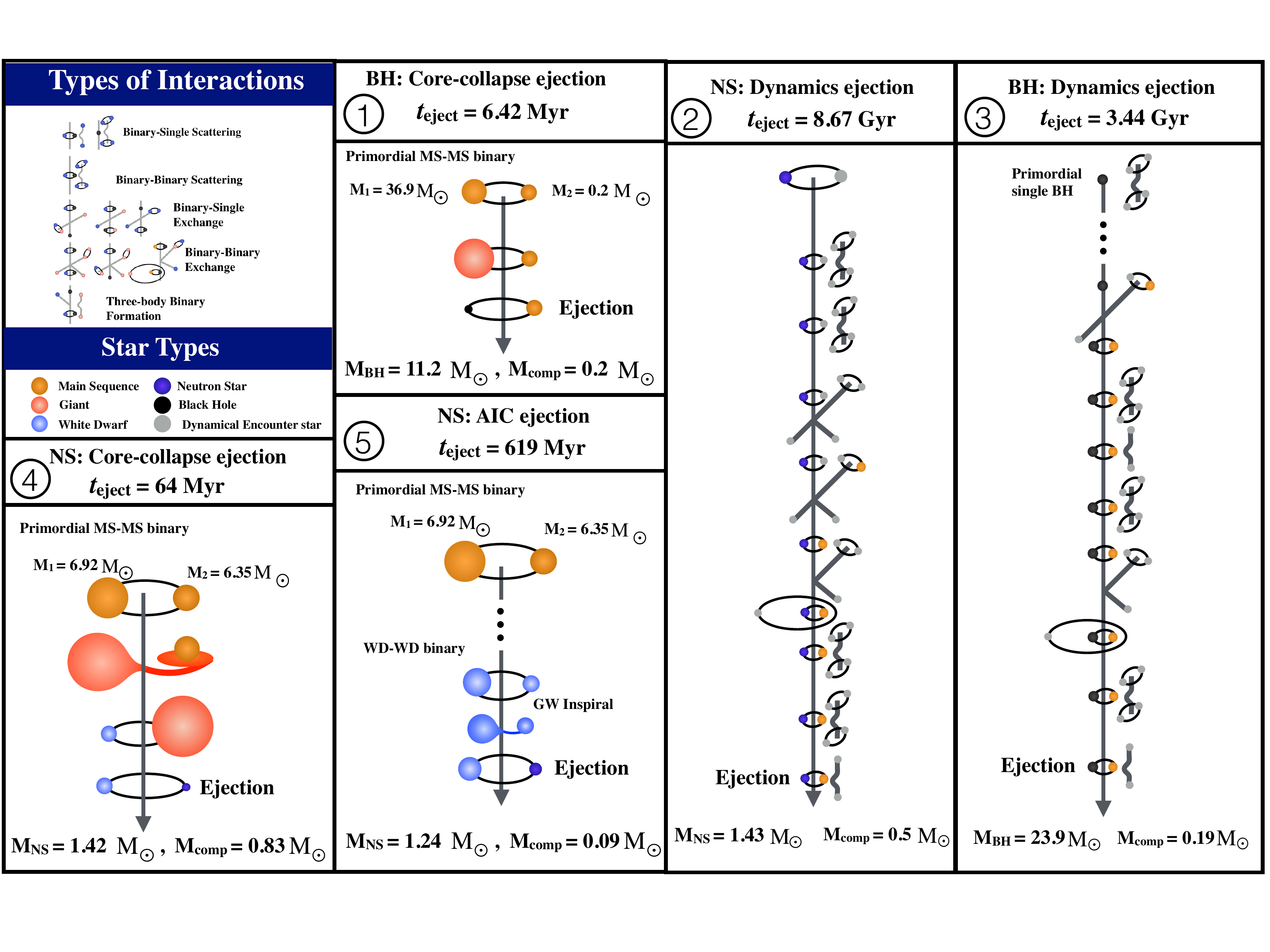}
\caption{\label{fig:cartoon} Illustration of the evolution of five specific systems which characterize the different systems shown in Figure \ref{fig:ejection}. Panel 1 shows the evolution of a BH--MS binary which is ejected through a CC SN. Panel 2 shows the evolution of a NS--MS binary which is ejected through a dynamical interaction, in this case a binary--single scattering event. Panel 3 shows the evolution of a BH--MS binary, also ejected through a binary--single scattering event. Panel 4 shows the evolution of a NS--WD binary which is ejected by the CC SN associated with the formation of the NS. Finally, panel 5 illustrates the evolution of a NS--WD binary which is ejected by AIC. The various star types and dynamical interaction types are denoted in the guide. }
\end{center}
\end{figure*}

We refer to all ejected BH and NS binaries that will eventually mass transfer as BH--MTBs and NS--MTBs, respectively. Ejected BH/NS binaries that are ejected but never mass transfer are not considered in this analysis.

Figure \ref{fig:ejection} shows the cumulative distribution of all ejected NS--MTBs (top two panels) and BH--MTBs (bottom two panels) versus the time of ejection, from their host cluster. The background gray histograms show the total number of  NS--MTBs and BH--MTBs ejected. As shown, a total of 288 NS--MTBs and 175 BH--MTBs are ejected from all models by 12 Gyr.

As discussed in Section \ref{sec:Ejection_mechanisms}, BH--MTBs and NS--MTBs can be ejected as the result of kicks imparted through dynamical interactions, CC SN, and AIC. The left-hand panels of Figure \ref{fig:ejection} show the contributions of these three distinct mechanisms to the total population of ejected MTBs. Black shows the systems ejected as the result of CC SN, red shows systems ejected through AIC, and blue shows systems ejected through dynamical encounters. As shown in the left-hand panels, NS--MTBs are ejected primarily through non-dynamical mechanisms (276 total from the CC SN and AIC channels versus 12 from dynamics channel) while BH--MTBs are ejected primarily through dynamical encounters (161 from dynamics channel versus 14 from the CC SN channel). This is to be expected for a variety of reasons. First, NS natal kicks are expected to be higher than BH natal kicks, leading to more frequent ejection of NSs from their cluster upon formation relative to the BH population. Second, BHs, being more massive than the typical object in a cluster, will rapidly mass-segregate upon formation, forming a dense sub-cluster where dynamical encounters with other objects become frequent. Eventually, these BHs are ejected from strong encounters with other objects. The number of dynamical encounters experienced by both BH-- and NS--MTBs is discussed further in Section \ref{sec:dyn_enc}.

%Note that depending upon the magnitude of the velocity kick imparted to each binary by its respective ejection mechanism, the binaries may be ejected promptly ($\lesssim 1$ Myr after SN/dynamical encounter), or may take as long as $\sim 1$ Gyr after after the kick to finally escape the host cluster's potential well. For example, as shown in the top right panel of Figure \ref{fig:ejection}, $\sim$ 20 binaries ejected by a CC SN (black curve) are shown to have $t_{\rm{eject}} > 1$ Gyr, long after the CC SN which created the NS, which typically occurs in the first 10--100 Myr of evolution.

In a handful of cases for non-dynamical ejections, binaries are not ejected from the cluster promptly after receiving the natal kick. At the time of birth of the BH/NS, the binary may still be bound albeit with large apocenter distances in the cluster-centric orbit. These binaries spend most of the remaining time without interacting with anything else at the low density outer regions of the cluster. Over time, as the cluster tidally loses mass, and the tidal boundary shrinks, the binary may be lost from the cluster. For example, as shown in the top right panel of Figure \ref{fig:ejection}, $\sim$ 20 binaries ejected by a CC SN (black curve) are shown to have $t_{\rm{eject}} > 1$ Gyr, long after the CC SN which created the NS, which typically occurs in the first 10--100 Myr of evolution.

The right-hand panels of Figure \ref{fig:ejection} show the contributions of different donor types to the total numbers of ejected BH-- and NS--MTBs. The blue histograms show systems with a WD donor, orange histograms show systems with a main sequence (MS) donor, and red histograms show systems with giant donors. As shown by the top-left panel, the majority of NS--MTBs have WD donors (276 of 288 total systems), while 11 and 1 ejected systems have MS and giant donors, respectively. MS donors dominate over WD donors at early times ($0-50$ Myr) simply because at these early times many WD progenitors have yet to evolve off the main-sequence. Comparison of the top-left and top-right plots suggest that these NS-MS binaries ejected at early times are simply primordial binaries where the more massive component collapses into a NS and the associated CC SN ejects this newly formed NS--MS from the cluster.

As shown by the bottom-right panel of Figure \ref{fig:ejection}, nearly all BH--MTBs have MS donors, with the exception of one system, ejected at $\simeq 6.5$ Gyr, which has a WD donor.

The circled numbers, 1--5, in Figure \ref{fig:ejection} each mark a specific model system. The evolution of each of these 5 systems is illustrated in Figure \ref{fig:cartoon}. %Panel 1 of Figure \ref{fig:cartoon} shows the evolution of a BH--MS binary which is ejected through a CC SN. Panel 2 shows the evolution of a NS--MS binary which is ejected through a dynamical interaction, in this case a binary--single scattering event. Panel 3 shows the evolution of a BH--MS binary, also ejected through a binary--single scattering event. Panel 4 shows the evolution of a NS--WD binary which is ejected by the CC SN associated with the formation of the NS. Finally, panel 5 illustrates the evolution of a NS--WD binary which is ejected by AIC.
All five panels of Figure \ref{fig:cartoon} note the time of ejection from the host cluster as well as the masses of the binary components at the time of ejection.

\subsection{Properties of BH--MTBs}
\begin{figure}
\begin{center}
\includegraphics[width=0.45\textwidth]{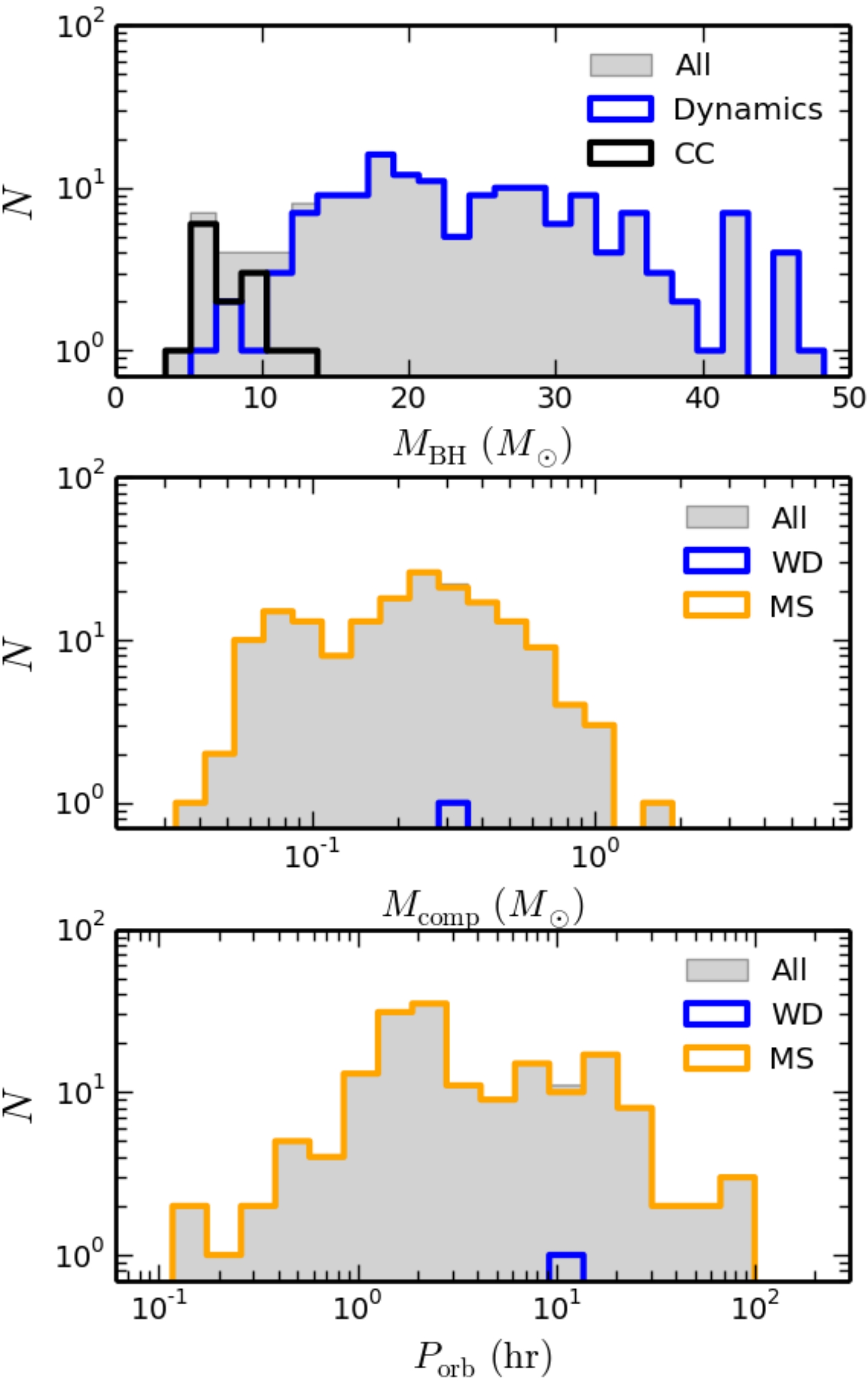}
\caption{\label{fig:BHparams} Binary parameters at time of ejection for all BH--MTBs ejected from our GC models. The top panel shows the distribution of BH masses, the middle panel shows the distribution of donor masses, and the bottom panel shows the distribution of orbital periods. The background gray histograms show the total values for all BH--MTBs, and the different colored plots show the breakdown into various ejection channels (top panel) and the various donor types (middle and bottom panels). The colors here are the same as in Figure \ref{fig:ejection}.}
\end{center}
\end{figure}

Figure \ref{fig:BHparams} shows the properties for all ejected BH--MTBs at the time of ejection. The top panel shows the distribution of BH masses, the middle panel shows the distribution of donor masses, and the bottom panel shows the distribution of orbital periods. As in Figure \ref{fig:ejection}, the background gray histograms show the total values for all BH--MTBs, and the different colored plots show the breakdown into various ejection channels (top panel) and the various donor types (middle and bottom panels). 

As the top panel shows, unlike the BH--MTBs ejected through dynamical interactions (blue), which span the entire BH mass spectrum, the BH--MTBs ejected through the core-collapse mechanism (black) are preferentially lower mass. This is as expected. The fallback prescription for BH natal kicks utilized within \texttt{CMC} supplies fallback proportional to the mass of the BH's progenitor. The progenitors of the least massive BHs will experience the least amount of fallback, and therefore will receive preferentially higher natal kicks. 

As the middle panel of Figure \ref{fig:BHparams} shows, the companion masses of the ejected BH--MTBs at the time of ejection are low ($\lesssim 1\,M_{\odot}$), consistent with systems that would be identified as LMXBs. The companion masses shown here are also consistent with companion masses predicted for BH--MTBs retained in their host cluster at late times, as discussed in \citet{Kremer2017}.

As the bottom panel of Figure \ref{fig:BHparams} shows, the orbital periods of the BH--MTBs at the time of ejection span several orders of magnitude from several minutes up to several days. Note that the $P_{\rm{orb}}$ values shown here are at the time of ejection, not at the time of mass-transfer onset. Depending upon the value of $P_{\rm{orb}}$ at the time of ejection, these systems will take varying amounts of time to inspiral to the onset of Roche-lobe overflow (through tidal decay and/or GR effects). The time of mass-transfer onset for all ejected MTBs is discussed further in Section \ref{sec:xray_present}.

\subsection{Properties of NS-MTBs}
\begin{figure}
\begin{center}
\includegraphics[width=0.45\textwidth]{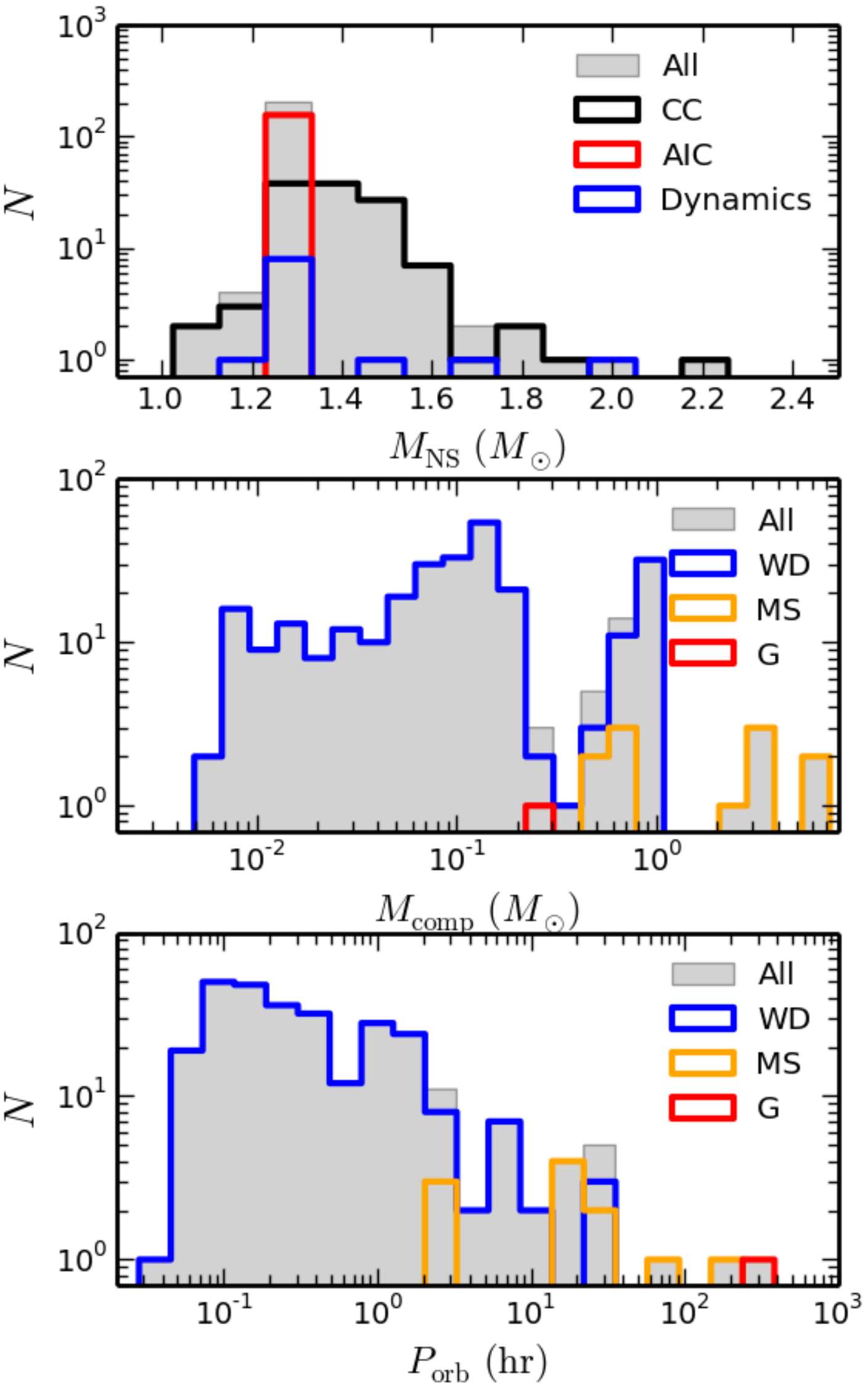}
\caption{\label{fig:NSparams} Same as Figure \ref{fig:BHparams} but for all NS--MTBs ejected from our GC models.}
\end{center}
\end{figure}
Figure \ref{fig:NSparams} is analogous to Figure \ref{fig:BHparams} but for all ejected NS--MTBs.

As the shown in the top panel of Figure \ref{fig:NSparams} shows, the mass spectrum of NSs that are members of the ejected NS--MTBs features a prominent peak at $1.24\,M_{\odot}$ as the result of NSs ejected through the AIC channel. This  value corresponds to the fixed NS mass assigned in \texttt{BSE} to all NSs formed through the AIC channel. The NS--MTBs ejected through the CC SN and dynamics channels span a wider spectrum in neutron mass.

Unlike the BH--MTBs, which almost exclusively have MS donors, NS--MTBs have significant numbers of both MS and WD donors (plus one NS--MTB ejected with a giant donor). As the middle panel of Figure \ref{fig:NSparams} shows, the majority of WD donor are very low mass ($\lesssim 0.01\,M_{\odot}$). This is a direct result of the standard evolutionary path through which these NS--WD binaries are formed. As panel 5 of Figure \ref{fig:cartoon} shows, NSs formed through AIC must accrete matter from their companion star to be pushed over the Chandrasekhar limit. In this manner, the donor star can be depleted of a significant portion of its mass, leaving behind an ultra-low mass WD.

As the orange plot of the middle panel of Figure \ref{fig:NSparams} shows, MS donors typically have relatively higher mass than the WD donors. This is because NS--MS binaries are typically ejected as detached systems, in contrast to the NS--WD binaries, many of which are ejected in a mass-transferring configuration, as discussed in previous paragraph. The range in donor masses for the NS--MTBs is also reflected in the wide range in $P_{\rm{orb}}$, as shown in the bottom panel of Figure \ref{fig:NSparams}. NS--WD binaries ejected in a mass-transferring configuration will have $P_{\rm{orb}} \lesssim 1$ hour, consistent with observed values of such systems \citep[e.g.,][]{intZand2005, intZand2007}. In contrast, the NS--MS binaries ejected as detached systems with $P_{\rm{orb}}$ values of \~ days will start mass-transferring only after the orbital separation of the now isolated ejected system shrinks as the result of standard binary star evolution, such as tidal interactions. %The time of mass-transfer onset is for all ejected MTBs is discuss further in Section \ref{sec:xray_present}.

\subsection{Dynamical Encounters}
\label{sec:dyn_enc}

\begin{figure}
\begin{center}
\plotone{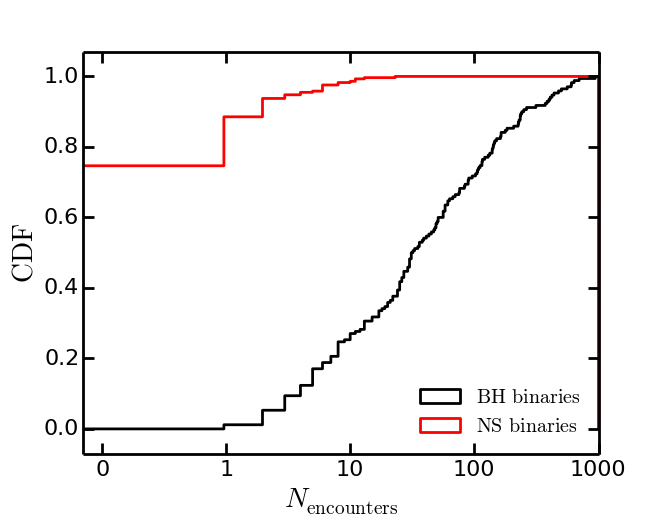}
\caption{\label{fig:Nenc} CDF of the number of dynamical encounters experienced by the accretor of each MTB before ejection from host cluster. The red plot shows the number of encounters for NSs and black shows BHs.}
\end{center}
\end{figure}
Figure \ref{fig:Nenc} shows the cumulative distribution function (CDF) of the number of dynamical encounters experienced by the accretor of each ejected MTB prior to ejection. The red plot shows the distribution for NSs and the black plot shows BHs. 73 of the 288 NSs experience at least one dynamical encounter before ejection. The remaining  215 (approximately 75\% of all ejected NS--MTBs) are ejected at the time of birth before any dynamical encounters.  173 of the 175 BHs experience at least one dynamical encounter before ejection. The BHs experience a median of 32 dynamical encounters before ejection.

As expected, dynamical interactions play a more significant role in the evolution and ejection of BH--MTBs than for NS--MTBs. This is expected for two reasons: First, unlike BHs, which are generally retained at birth, many NSs are ejected at birth due to natal kicks before ever undergoing a dynamical encounter. Second, even the NSs that are retained after birth (e.g., those formed through the AIC channel which receive low velocity kicks) do not reside in the densest part of the cluster center, having been pushed out by the BHs. Thus, the number of dynamical encounters these retained NSs undergo, on average, is low, relative to their BH counterparts.
% * <sitari@gmail.com> 2017-12-16T19:43:51.215Z:
% 
% > As expected, dynamical interactions play a more significant role in the evolution and ejection of BH--MTBs than for NS--MTBs. This is simply because BHs mass-segregate to the more dense central regions of their host cluster where the rate of dynamical interactions is higher.
% I think this is simply because most BHs are retained after birth as opposed to most NSs are ejected at birth. What you really mean is that even the NSs that are retained after birth (or AIC, ECSN NSs) don't reside at the densest part of the cluster center being pushed out by the BHs. Thus number of dynamical encounters they go through is lower. In addition, it is easier to dynamical eject the lower-mass NSs from a shallower part of the cluster potential, and thus, a lower number of dynamical encounters are needed to sufficiently harden and eject them. 
% 
% ^ <kylekremer23@gmail.com> 2017-12-17T00:37:47.087Z.

Although the majority of our NS--MTBs experience no dynamical encounters once formed, dynamical interactions do play an important role in the formation of some of our model NS--MTBs. Of the 288 total NS--MTBs, 251 are primordial binaries, meaning the remaining 37 where formed as the result of dynamical interactions.

\section{Modeling orbits of ejected systems in Galactic potential}
\label{sec:gx-ev}
\begin{figure}
\begin{center}
\plotone{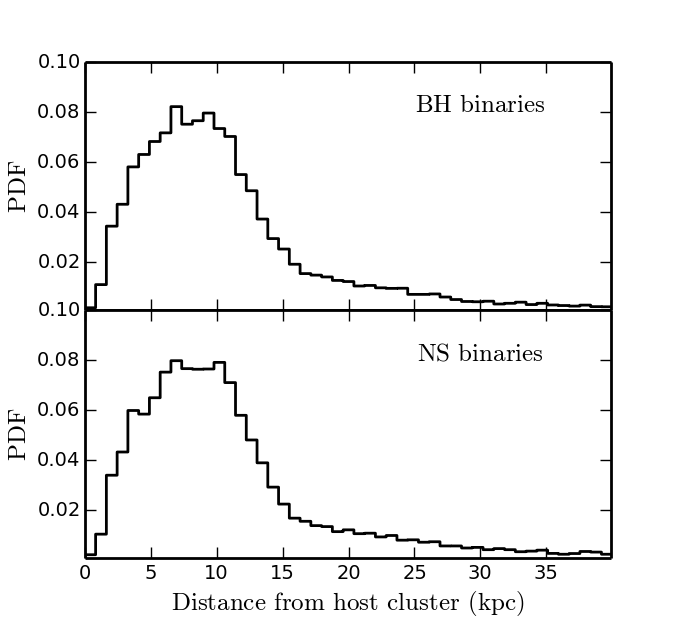}
\caption{\label{fig:distance} Distribution of the distance each ejected MTB has traveled from its host cluster by present day. The top panel shows the distribution for BH--MTBs and the bottom panel, NS--MTBs.}
\end{center}
\end{figure}

\begin{figure*}
\begin{center}
\includegraphics[width=1.0\linewidth]{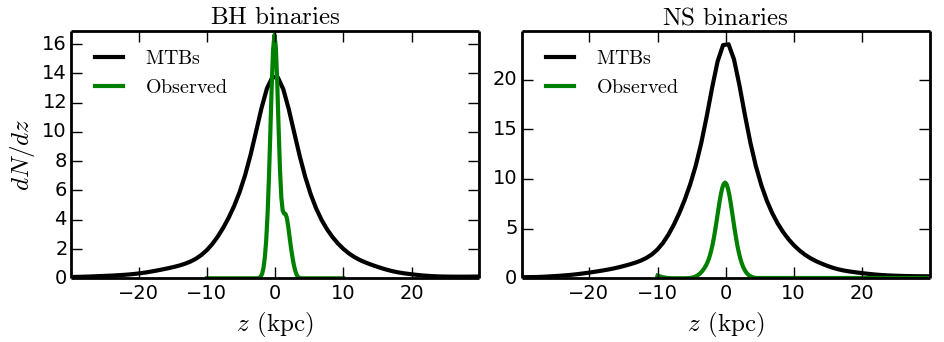}
\caption{\label{fig:z_dist} Black curves shows the number density of the $z$-position of all the 175 total model BH--MTBs (left panel) and the 288 total NS--MTBs (right panel) at the present day. Green curves show number densities of $z$-positions of all observed XRBs from \citet{CorralSantana2016} (BH--XRBs) and \citet{Jonker2004} (NS--XRBs).}
\end{center}
\end{figure*}

The calculations performed within \texttt{CMC} determine the properties of each MTB at the time of ejection from the host cluster. In order to determine the present-day spatial distribution of these binaries within the Galactic field, we must continue to model the orbits of these systems within the Galactic potential after they are ejected.

We use the \texttt{GalPot} code, which provides routines to calculate orbits of objects in the Milky Way potential, using the potentials found in \citet{McMillan2017} and \citet{Dehnen1998}. \texttt{GalPot} takes as input the initial velocity and position of an object and determines the complete orbit over any specified duration in time.

In order to fully specify the initial position and velocity of each ejected MTB within the galaxy, we must first select a position and velocity for each ejected MTB's host cluster at the time of ejection. We draw initial cluster positions from the $x,y,z$ positions of all observed MW GCs given in \citet{Harris1996}. Initial velocities of each host cluster are then determined by simply assuming a circular Kepler orbit, $v=\sqrt{G M_{\rm{enc}}/r}$, where $M_{\rm{enc}}$ is the total mass enclosed within the galactocentric radius (also drawn from observed values from \citet{Harris1996}). $M_{\rm{enc}}$ is found using the Galactic mass densities specified in \citet{McMillan2017}. This velocity is then broken down into its components, by randomly selecting an initial $\dot{\theta}$ and $\dot{\phi}$ then converting to cylindrical coordinates ($\dot{R}$, $\dot{z}$, and $\dot{\phi}$) using the same  $\theta$ and $\phi$ values used to determine the initial position. Note that, for simplicity, we assume all GC orbits are circular ($\dot{r} = 0$).

Given initial values for $R$, $z$, $\phi$, $\dot{R}$, $\dot{z}$, and $\dot{\phi}$ for the host cluster of each ejected MTB, the position and velocity of each host cluster can be determined for any later time using \texttt{GalPot}. For each ejected MTB, we integrate its host cluster forward in time until $t_{\rm{eject}}$, the time that the MTB is ejected, in order to determine $R$, $z$, $\phi$, $\dot{R}$, $\dot{z}$, and $\dot{\phi}$ of the host cluster at the time of ejection.

The magnitude of the velocity after ejection, $v_{\rm{eject}}$, and time of ejection, $t_{\rm{eject}}$, for each ejected system is determined by \texttt{CMC}. We randomly select an ejection angle for each system (by assuming any direction is equally probable) in order to determine the components of $v_{\rm{eject}}$ in the cluster frame. We then combine these components with $\dot{R}$, $\dot{z}$, and $\dot{\phi}$ for the host cluster at $t_{\rm{eject}}$ to determine $\dot{R}_{\rm{MTB}}$, $\dot{z}_{\rm{MTB}}$, and $\dot{\phi}_{\rm{MTB}}$ for each ejected MTB in the galactocentric frame. The position of the ejected MTB at $t_{\rm{eject}}$ is simply given by the position of the host cluster at $t_{\rm{eject}}$.

Given $R_{\rm{MTB}}$, $z_{\rm{MTB}}$, $\phi_{\rm{MTB}}$, $\dot{R}_{\rm{MTB}}$, $\dot{z}_{\rm{MTB}}$, and $\dot{\phi}_{\rm{MTB}}$ at $t_{\rm{eject}}$, each ejected MTB can then be integrated forward in time to the present day in order to determine its current location. Additionally, each host cluster can also be integrated forward to the present day to determine the present-day distance each MTB is from its host cluster. The time at present day is determined by the age of each host cluster. In order to reflect the uncertainty in GC ages, we consider here 7 equally probable cluster ages in the range $9-12$ Gyr, spaced 0.5 Gyr apart.

This entire process (randomly selecting an initial position and velocity for each MTB's host cluster, integrating the host cluster forward to $t_{\rm{eject}}$, randomly selecting an ejection angle for each ejected system, integrating the ejected system and host cluster forward in time from $t_{\rm{eject}}$ to present day) is repeated 140 times for each ejected MTB, in order generate a larger sample.

Figure \ref{fig:distance} shows the distribution of the distance each ejected MTB has traveled from its host cluster by the present day. The top panel shows the distribution for BH--MTBs and the bottom panel, NS--MTBs. As shown in this figure, our model ejected systems have traveled several kpc or more from their host cluster by the time they would potentially be observed as X-ray binaries (XRBs). This suggests it is unlikely any observed XRB which originated from within a GC would be spatially associated with its host cluster at the time of observation.

\subsection{Comparison to spatial distribution of observed LMXBs}
\label{sec:spatial_dist}

Figure \ref{fig:z_dist} shows the number densities (generated using kernel density estimation) of present day $z$-positions of our ejected MTBs compared to observed Galactic XRBs with known $z$-positions. Black curves shows the number density of the $z$-position of all the 175 total model BH--MTBs (left panel) and the 288 total NS--MTBs (right panel) at the present day.  Green curves show number densities of observed XRBs with published $z$-positions from \citet{CorralSantana2016} (BH--XRBs) and \citet{Jonker2004} (NS--MTBs).

As shown in Figure \ref{fig:z_dist}, while the observed XRB systems are sharply peaked in the Galactic plane ($z=0$), the model systems are spherically distributed, reflecting the observed distribution of GCs in the galaxy.

\section{Determining X-ray observability}
\label{sec:luminosity}

\begin{figure*}
\begin{center}
\includegraphics[width=0.8\textwidth]{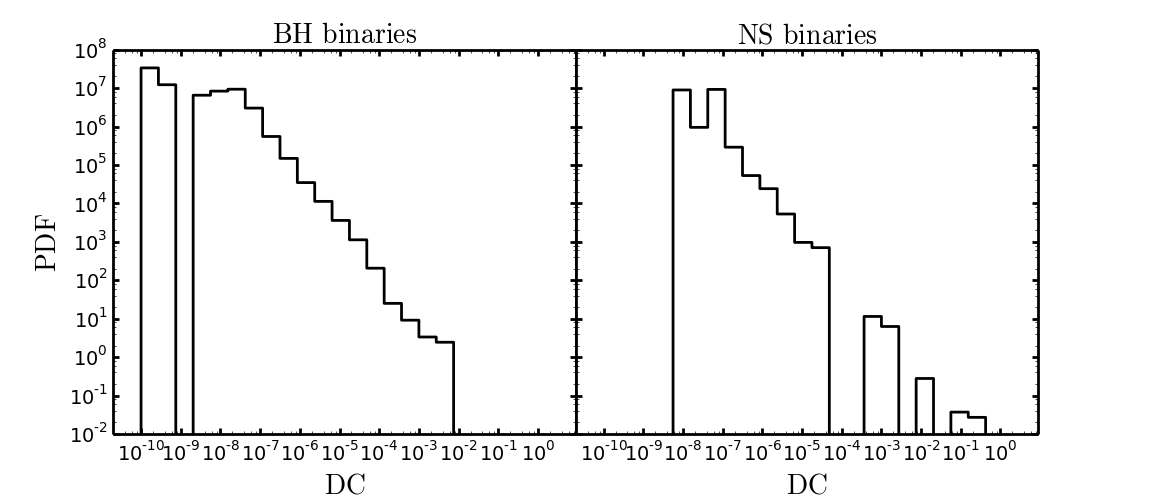}
\caption{\label{fig:DC_12gyr} Duty cycles for all BH--MTBs (left-panel) and NS--MTBS (right panel) calculated as described in Section \ref{sec:luminosity}.
}
\end{center}
\end{figure*}

As ejected MTBs are evolved as isolated binary systems following ejection from their host clusters, we continue to track all binary properties, including mass-transfer rates as function of time. The mass-transfer rate of each MTB at the time of observation, as well as the type of donor, determine whether or not each system could be observed as an X-ray source. We follow a similar method to that of \citet{Fragos2008} to determine the X-ray luminosities and duty cycles of the sources.

The ratio of a binary's mass-transfer rate to the critical mass-transfer rate for irradiated accretion disks,  $\dot{M}_{\rm{crit}}$, determines whether the system is observed as a persistent source or a transient source. As in \citet{vanHaaften2015}, we use the equation for $\dot{M}_{\rm{crit}}$ derived by \citet{Dubus1999}, in the form given by \citet{intZand2007}:

\begin{equation}
\label{eq:Mdot_crit}
\dot{M}_{\rm{crit}} = 5.3 \times 10^{-11} \, f \, \left( \frac{M_A}{M_{\odot}} \right)^{0.3} \left( \frac{P_{\rm{orb}}}{\rm{hr}} \right)^{1.4} M_{\odot} \, \rm{yr}^{-1}
\end{equation}
Here, $M_A$ is the accretor mass, $P_{\rm{orb}}$ is the binary orbital period, and $f$ is a scale factor depending on the disk composition. We use $f =1$ for solar-composition disks (MS donors) and $f=6$ for helium disks (WD donors).

Binaries with mass-transfer rates higher than $\dot{M}_{\rm{crit}}$ are expected to be persistent sources. For such systems the X-ray luminosity is determined by

\begin{equation}
\label{eq:Lx}
L_{X, \, \rm{persistent}}  = \eta_{\rm{bol}}\, \epsilon \, \frac{G M_A \dot{M}_D}{R_A}
\end{equation}

where $R_A$ is the accretor radius and $\dot{M}_D$ is the mass-transfer rate. As in \citet{Fragos2008}, we use $R_A=10$ km for NS accretors and $R_A=3 R_s$, where $R_s$ is the Schwarzschild radius ($R_s = 2G M_{\rm{BH}}/c^2$), for BH accretors. $\eta_{\rm{bol}}$ is a factor that converts the bolometric luminosity to the X-ray luminosity in the Chandra energy band ($0.3-8 $ keV). As in \citet{Fragos2008}, for accreting BHs this conversion factor is estimated to be $\eta_{\rm{bol}}=0.8$ \citep{Miller2001} and for accreting NSs, $\eta_{\rm{bol}}=0.55$ \citep{DiSalvo2002,Maccarone2003,PortegiesZwart2004}. $\epsilon$ is the conversion efficiency of gravitational binding energy to radiation. For NS accretors (for which the mass-transfer stream impacts the surface directly), we use $\epsilon = 1.0$ and for BH accretors (for which the mass-transfer stream impacts the disk), we use $\epsilon = 0.5$

Binaries with mass-transfer rates lower than $\dot{M}_{\rm{crit}}$ are expected to be transient sources. A  transient source will alternate between periods of outburst (when they are observable as X-ray sources) and quiescence (when they are too faint to be detectable). The fraction of time these transient systems spend in outburst defines the duty cycle ($\rm{DC}$):

\begin{equation}
\rm{DC} = \frac{T_{\rm{outburst}}}{T_{\rm{outburst}} + T_{\rm{quiescence}}}
\end{equation}
The details of the thermal disk instability model are poorly understood. As noted in \citet{Fragos2008}, a simple, physically motivated treatment is to assume that for transient sources in the quiescent state, matter from the donor is accumulated in the disk and no matter is accreted onto the accretor. In the outburst state all of this matter is accreted onto the accretor, emptying the disk. Taking into account, as in \citet{Fragos2008}, that the X-ray luminosity probably cannot exceed the Eddington luminosity, $L_{\rm{Edd}}$, by more than a factor of 2 \citep[see][]{Taam1997}, we can define the X-ray luminosity of transient systems during outburst as:

\begin{equation}
\label{eq:Lx-transient}
L_{X, \, \rm{transient}} = \eta_{\rm{bol}} \epsilon \times \rm{min}\Big[ 2\times \it{L}_{\rm{Edd}}, \left( \frac{\it{G} M_A \dot{M}_D}{R_A} \frac{\rm{1}}{\rm{DC}} \right) \Big]
\end{equation}

As noted in \citet{Fragos2008} (which in turn uses the results of \citet{Dobrotka2006}), the general behavior of $\rm{DC}$ can be expressed in terms of known parameters as

\begin{equation}
\label{eq:DC}
\rm{DC} = \left( \frac{\dot{\it{M}}_D}{\dot{\it{M}}_{\rm{crit}}} \right)^2
\end{equation}

 allowing us to re-write Equation (\ref{eq:Lx-transient}) as 
 
 \begin{equation}
L_{X, \, \rm{transient}} = \eta_{\rm{bol}} \epsilon \times \rm{min}\Big[ 2\times \it{L}_{\rm{Edd}}, \left( \frac{\it{G} M_A \dot{M}_{\rm{crit}}^2}{\it{R}_A \dot{M}_D} \right) \Big]
\end{equation}
 Given the orbital parameters, including the mass-transfer rate, of our ejected MTBs at the present day, we use this method to determine whether or not each system is persistent or transient, in the case of transient systems, determine the duty cycle, and calculate X-ray luminosities.

\begin{figure}
\begin{center}
\plotone{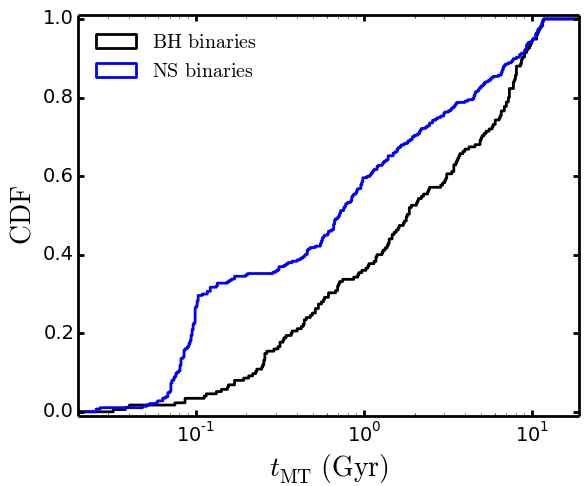}
\caption{\label{fig:t_MT} Cumulative distribution function of time of mass-transfer onset for all BH--MTBs (black) and NS--MTBs (blue).}
\end{center}
\end{figure}

\subsection{X-ray features of ejected systems at present day}
\label{sec:xray_present}

\begin{deluxetable*}{ccc|c|cc|c|c|c|c}
\tabletypesize{\scriptsize}
\tablewidth{0pt}
\tablecaption{BH--XRB orbital parameters. \label{table:BH-XRBs}}
\tablehead{
	\colhead{$t_{\rm{obs}}$} &
    \colhead{$t_{\rm{eject}}$} &
    \colhead{$t_{\rm{MT}}$} &
    \colhead{Comp. type} &
    \colhead{$M_{\rm{BH}}$} &
    \colhead{$M_{\rm{comp}}$} &
    \colhead{$a$} &
    \colhead{$\dot{M}_D$} &
    \colhead{$\rm{DC}$}&
    \colhead{$L_X$} \\
    \multicolumn{3}{c}{(Gyr)} &
    \colhead{} &
    \multicolumn{2}{c}{($M_{\odot}$)} &
    \colhead{(AU)} &
    \colhead{($M_{\odot}\,\rm{yr}^{-1}$)} &
    \colhead{} &
    \colhead{($\rm{erg\, s}^{-1}$)}
}
\startdata
9.0 & 7.89 & 7.89 & MS & 14.17 & 0.05 & 0.0095 & $4.39 \times 10^{-11}$ & 0.02 & $7.87 \times 10^{37}$\\
9.0 & 8.73 &  8.73 & MS & 20.38 & 0.05 & 0.01 & $4.47 \times 10^{-10}$ & 0.01 & $1.12 \times 10^{38}$\\
9.0 & 8.78 &  8.80 & MS & 13.88 & 0.35 & 0.01 & $1.01 \times 10^{-9}$ & 1.0 & $1.15 \times 10^{37}$\\
9.0 & 8.72 &  8.72 & MS & 13.33 & 0.21 & 0.0076 & $7.12 \times 10^{-10}$ & 1.0 & $8.10 \times 10^{36}$\\
\hline
9.5 & 3.89 & 9.32 & MS & 11.62 & 0.33 & 0.0092 & $7.59 \times 10^{-10}$& 1.0 & $8.63 \times 10^{36}$\\
9.5 & 8.78 & 8.81 & MS & 14.18 & 0.05 & 0.0093 & $5.17 \times 10^{-11}$ & 0.02 & $5.94 \times 10^{37}$\\
9.5 & 9.1 & 9.1 & MS & 13.65 & 0.45 & 0.013 & $5.66 \times 10^{-10}$ & 0.82 & $1.95 \times 10^{37}$\\
\hline
10.0 & $6.6 \times 10^{-3}$ & 9.82 & MS & 10.11 & 0.11 & 0.0044 & $1.62 \times 10^{-9}$ & 1.0 & $1.84 \times 10^{37}$\\
10.0 & 3.89 & 9.32 & MS & 11.9 & 0.05 & 0.0086 & $5.14 \times 10^{-11}$ & 0.03 & $4.91 \times 10^{37}$\\
10.0 & $8.1 \times 10^{-3}$ & 9.94 & MS & 6.68 & 0.1 & 0.0043 & $8.74 \times 10^{-10}$ & 1.0 & $9.94 \times 10^{36}$\\
10.0 & 9.1 & 9.1 & MS & 13.94 & 0.16 & 0.007 & $5.92 \times 10^{-10}$ & 1.0 & $6.74 \times 10^{36}$\\
\hline
11.0 & 0.5 & 10.68 & MS & 18.28 & 0.05 & 0.0094 & $1.19 \times 10^{-10}$ & 0.15 & $2.20 \times 10^{37}$\\
11.0 & 10.8 & 10.83 & MS & 17.52 & 0.05 & 0.0099 & $5.88 \times 10^{-11}$ & 0.03 & $5.68 \times 10^{37}$\\
11.0 & 10.75 & 10.75 & MS & 14.72 & 0.26 & 0.0087 & $7.68 \times 10^{-10}$ & 1.0 & $8.74 \times 10^{36}$\\
\hline
11.5 & $6.8 \times 10^{-3}$ & 11.09 & MS & 9.56 & 0.05 & 0.0084 & $3.17 \times 10^{-11}$ & 0.01 & $8.92 \times 10^{37}$\\
\enddata
\tablecomments{Orbital parameters for all BH--MTBs with $\rm{DC} \geq 0.01$ measured at all seven observations times from $9-12$ Gyr. $t_{\rm{obs}}$ denotes the time of observation and $t_{\rm{MT}}$ denotes the time mass-transfer begins for each binary.}
\end{deluxetable*}

First, we assume, as in Section \ref{sec:gx-ev}, that all clusters are 12 Gyr old and integrate all ejected MTBs from $t_{\rm{eject}}$ to 12 Gyr to determine the orbital parameters of the binaries at the present day.

Figure \ref{fig:DC_12gyr} shows the distribution of $\rm{DC}$ for all binaries at 12 Gyr. The left-hand panel shows the distribution for BH--MTBs and the right-hand panel, NS--MTBs. At 12 Gyr, none of the ejected sysems are observed as persistent X-ray sources. Only 4 of the 288 NS--MTBs and 0 of the 175 BH--MTBs have a $\rm{DC} \geq 0.01$.

These low duty cycles are primarily a consequence of the fact that most systems start mass-transferring a few Gyr or less after the birth of their host cluster, which means by late times, when these systems would potentially be observed, the mass-transfer rates have fallen significantly. Figure \ref{fig:t_MT} shows the CDF of the time of mass-transfer onset for the ejected MTBs. BH--MTBs and shown on the left and NS--MTBs on the right. 36\% (63 out of 175) of all BH--MTBs and 59\% (171 out of 288) of all NS--MTBs begin mass transfer before $t=1$ Gyr. 

In order to generate a larger sample of X-ray observable systems, we calculate the $\rm{DC}$ of each system at seven different observations between 9 and 12 Gyr, spaced 0.5 Gyr apart, as in Section \ref{sec:gx-ev}. Tables \ref{table:BH-XRBs} and \ref{table:NS-XRBs} show the orbital parameters for all BH--XRBs and NS--XRBs, respectively, which are observed with $\rm{DC} \geq 0.01$ for any of the six observation times. Note that $\rm{DC} = 0.01$ is chosen as an arbitrary lower limit in order to select a small set of characteristic systems for which we can show orbital parameters.

\begin{deluxetable*}{ccc|c|cc|c|c|c|c}
\tabletypesize{\scriptsize}
\tablewidth{0pt}
\tablecaption{NS--XRB orbital parameters. \label{table:NS-XRBs}}
\tablehead{
	\colhead{$t_{\rm{obs}}$} &
    \colhead{$t_{\rm{eject}}$} &
    \colhead{$t_{\rm{MT}}$} &
    \colhead{Comp. type} &
    \colhead{$M_{\rm{NS}}$} &
    \colhead{$M_{\rm{comp}}$} &
    \colhead{$a$} &
    \colhead{$\dot{M}_D$} &
    \colhead{$\rm{DC}$}&
    \colhead{$L_X$} \\
    \multicolumn{3}{c}{(Gyr)} &
    \colhead{} &
    \multicolumn{2}{c}{($M_{\odot}$)} &
    \colhead{(AU)} &
    \colhead{($M_{\odot}\,\rm{yr}^{-1}$)} &
    \colhead{} &
    \colhead{($\rm{erg\, s}^{-1}$)}
}
\startdata
9.0 & 8.67 & 8.69 & MS & 1.65 & 0.28 & 0.0052 & $1.27 \times 10^{-10}$ & 0.38 & $4.61 \times 10^{36}$\\
\hline
9.5 & 8.67 & 8.69 & MS & 1.72 & 0.22 & 0.0047 & $1.17 \times 10^{-10}$ & 0.5 & $3.39 \times 10^{36}$\\
9.5 & 9.43 & 9.43 & WD & 1.31 & 0.02 & 0.0016 & $7.38 \times 10^{-11}$ & 0.33 & $2.47 \times 10^{36}$\\
9.5 & 9.17 & 9.31 & WD & 1.31 & 0.02 & 0.0019 & $2.82 \times 10^{-11}$ & 0.02 & $1.35 \times 10^{37}$\\
9.5 & 9.39 & 9.39 & WD & 1.31 & 0.02 & 0.0017 & $5.02 \times 10^{-11}$ & 0.11 & $4.88 \times 10^{36}$\\
\hline
10.0 & 8.68 & 8.69 & MS & 1.77 & 0.16 & 0.004 & $1.13 \times 10^{-10}$ & 0.89 & $1.88 \times 10^{36}$\\
\hline
10.5 & 8.68 & 8.69 & MS & 1.85 & 0.09 & 0.0034 & $1.60 \times 10^{-10}$ & 1.0 & $1.36 \times 10^{36}$\\
10.5 & 10.35 & 10.35 & WD & 1.3 & 0.02 & 0.0019 & $3.42 \times 10^{-11}$ & 0.04 & $9.48 \times 10^{36}$\\
\hline
11.0 & 8.68 &8.69 & MS & 1.87 & 0.06 & 0.0046 & $2.11 \times 10^{-11}$ & 0.02 & $2.00 \times 10^{37}$\\
11.0 & 10.87 & 10.89 & WD & 1.26 & 0.02 & 0.0018 & $4.25 \times 10^{-11}$ & 0.07 & $6.07 \times 10^{36}$\\
11.0 & 8.22 & 10.56 & MS & 1.54 & 0.29 & 0.0054 & $1.00 \times 10^{-10}$ & 0.19 & $6.83 \times 10^{36}$\\
11.0 & 10.78 & 10.78 & WD & 1.31 & 0.02 & 0.002 & $2.44 \times 10^{-11}$ & 0.02 & $1.74 \times 10^{37}$\\
11.0 & 10.83 & 10.83 & WD & 1.31 & 0.02 & 0.0019 & $3.12 \times 10^{-11}$ & 0.03 & $1.13 \times 10^{37}$\\
\hline
11.5 & 11.38 & 11.38 & MS & 1.42 & 0.42 & 0.0062 & $2.09 \times 10^{-9}$ & 1.0 & $1.38 \times 10^{37}$\\
11.5 & 11.46 & 11.46 & WD & 1.28 & 0.03 & 0.0014 & $1.37 \times 10^{-10}$ & 1.0 & $8.09 \times 10^{35}$\\
11.5 & 8.22 & 10.56 & MS & 1.58 & 0.24 & 0.0051 & $9.16 \times 10^{-11}$ & 0.21 & $5.81 \times 10^{36}$\\
11.5 & 11.46 & 11.46 & WD & 1.3 & 0.03 & 0.0014 & $1.46 \times 10^{-10}$ & 1.0 & $8.75 \times 10^{35}$\\
\hline
12.0 & 11.38 & 11.38 & MS & 1.57 & 0.27 & 0.0053 & $9.19 \times 10^{-11}$ & 0.17 & $6.98 \times 10^{36}$\\
12.0 &  11.88 & 11.88 & WD & 1.31 & 0.02 & 0.0018 & $4.37\times 10^{-11}$ & 0.08 & $6.24 \times 10^{36}$\\
12.0 & 0.02 & 11.68 & G & 2.21 & 0.28 & 0.0086 & $5.52 \times 10^{-11}$ & 0.01 & $1.01 \times 10^{38}$\\
12.0 & 8.22 & 10.56 & MS & 1.63 & 0.2 & 0.0046 & $8.17 \times 10^{-11}$ & 0.23 & $4.75 \times 10^{36}$\\
\enddata
\tablecomments{Same as Table \ref{table:BH-XRBs} but for NS--XRBs.}
\end{deluxetable*}

\subsection{Estimating the number of dynamically-formed LMXBs in the MW}
\label{sec:MW_LMXBs}

\begin{figure}
\begin{center}
\plotone{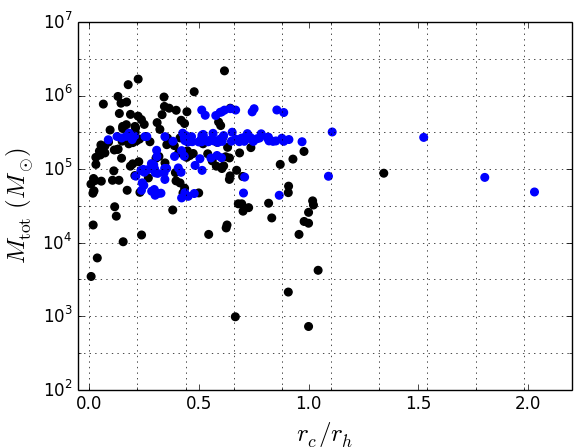}
\caption{\label{fig:weight1} Comparison of model GCs (blue) with observed MW GCs (black) in $r_c/r_h$--mass space. The grid lines show the bins used in weighting scheme.}
\end{center}
\end{figure}
\begin{figure}
\begin{center}
\plotone{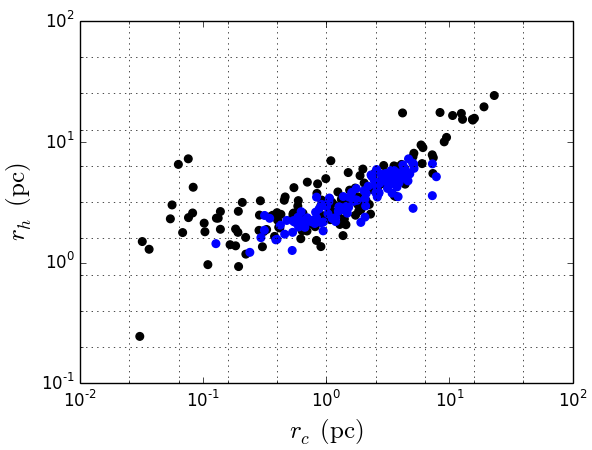}
\caption{\label{fig:weight2} Same as Figure \ref{fig:weight1}, but in $r_c-r_h$ plane.}
\end{center}
\end{figure}

In order to estimate the total number of dynamically-formed LMXBs in the MW at the present day, we use two different weighting schemes to weight our GC models according to how well they match observed MW GCs.

For the first weighting scheme, we use a method similar to \citet{Rodriguez2015} and compare our model GCs to the observed MW GCs (from \citet{Harris1996}) in the $r_c/r_h$--mass plane ($r_c$ and $r_h$ are the observed core radius and half-light radius, respectively). Figure \ref{fig:weight1} shows the comparison of our model GCs (blue) to observed MW clusters (black) in this scheme. Structural parameters for the model clusters are calculated at $t=12$ Gyr. To determine masses of observed GCs, we use absolute visual magnitudes from \citet{Harris1996} and obtain cluster mass by assuming a mass-to-light ratio of 2 and a constant bolometric correction of -0.107 (bolometric correction for the sun) for all models.

For the second weighting scheme, we compare our model GCs to the observed MW GCs in the $r_c$--$r_h$ plane. Figure \ref{fig:weight2} shows the comparison of our model GCs (blue) to observed MW clusters (black) in this scheme.

For each scheme, the model and observed clusters are binned into a $10 \times 10$ grid. Each model cluster is then given a weight according to: 
\begin{equation}
\rm{weight}_{\it{\,i}} = \frac{\it{N}_{\it{\,i}}^{\rm{\,obs}}}{\it{N}_{\it{\,i}}^{\rm{\,model}}}
\end{equation}
where $N_{\it{\,i}}^{\rm{\,obs}}$ and $N_{\it{\,i}}^{\rm{\,model}}$ are the number of observed clusters and model clusters in the $i^{\rm{th}}$ bin, respectively.

After applying each weighting scheme, we re-scale the total mass of all (weighted) GCs to match the total mass of the observed MW GC system, which is determined from \citet{Harris1996} to be $\sim 3.4 \times 10^7 M_{\odot}$.

These schemes are applied at each of the seven observation times between 9 and 12 Gyr discussed in Section \ref{sec:xray_present}. Assuming all observation times from 9--12 Gyr are equally likely, we estimate, using the first weighing scheme, 186 BH--MTBs and 328 NS--MTBs may exist at present within the MW that were formed in GCs. Using the methods of Section \ref{sec:luminosity} to place cuts on these totals based on the duty cycles of these systems at the time of observation, we estimate 1.0 and 5.4 of these systems will be observable at the present day as BH--XRBs and NS--XRBs, respectively.

Using the second weighting scheme, we estimate 181 BH--MTBs and 272 NS--MTBs may exist at present within the MW that were formed in GCs. After placing cuts on these totals based on the X-ray observability of systems, we estimate 1.2 and 4.2 of these systems will be observable at the present day as BH--XRBs and NS--XRBs, respectively.

To date, several hundred LMXBs have been observed in the MW \citep[e.g.,][]{Liu2007}. It has been shown that this observed population constitutes a small fraction of the total number of LMXBs in the galaxy, which may be as high as $\sim$  thousands \citep[e.g.,][]{vandenHeuvel2001, CorralSantana2016}. Therefore, we conclude that MTBs formed in and ejected from GCs are unlikely to contribute significantly to the population of field XRBs.

\section{Conclusion} \label{sec:conclusion}
We have explored the population of mass-transferring binaries ejected from globular clusters with both BH and NS accretors. We have demonstrated using a set of 137 globular cluster models of varying cluster properties that such systems can be ejected from clusters with white dwarf, main sequence, and giant donors and can have a wide range of orbital parameters. 

We calculated the orbits of these ejected systems in the Galactic potential to determine the present-day positions of these systems in the galaxy and compared to the distribution of observed XRBs in the Milky Way. We have shown that, as a result of the early ejection times and high escape velocites of these systems, most will be far (several kpc) from their host cluster by the time they are observed, meaning they are unlikely to be spatially associated with their host cluster.

Additionally, we have calculated the X-ray luminosities and duty cycles of these ejected mass-transferring binaries. We have shown that, as a result of the early mass-transfer-onset times for these binaries, most of these systems will have low duty cycles at the present day, and are unlikely to be observed as X-ray sources. We estimate $\sim 300$ mass-transferring NS binaries and $\sim 180$ mass-transferring BH binaries that formed dynamically in GCs may exist at present within the MW halo. Of these, we estimate, based on mass-transfer rates and duty cycles at the present time,  at most a few would observable as BH--XRBs and NS--XRBs. The total number of dynamically formed XRBs found in this study is unlikely to contribute significantly to the total number of low-mass X-ray binaries in the galaxy.

%%%%%%%%%%%%%%%%%%%%%%%%%%%%%%%%%3

\acknowledgments
This work was supported by NASA ATP Grant NNX14AP92G 
and NSF Grant AST-1716762. K.K. acknowledges support by the National Science Foundation Graduate Research Fellowship Program under Grant No. DGE-1324585.
S.C. acknowledges support from
CIERA, the National Aeronautics and Space Administration
through a Chandra Award Number TM5-16004X/NAS8-
03060 issued by the Chandra X-ray Observatory Center
(operated by the Smithsonian Astrophysical Observatory for and on behalf of the National Aeronautics
Space Administration under contract NAS8-03060), 
and Hubble Space Telescope Archival research 
grant HST-AR-14555.001-A (from the Space Telescope 
Science Institute, which is operated by the Association of Universities for Research in Astronomy, Incorporated, under NASA contract NAS5-26555).

\software{\texttt{CMC} \citep{Joshi2000,Joshi2001,Fregeau2003, Fregeau2007, Chatterjee2010,Chatterjee2013,Umbreit2012,Morscher2013,Rodriguez2016b}, \texttt{Fewbody} \citep{Fregeau2004}, \texttt{BSE} \citep{Hurley2002}, \texttt{GalPot} \citep{McMillan2017,Dehnen1998}}

\appendix

\startlongtable
\begin{deluxetable*}{lcccccc|cc|ccc|c|c||cc}
\tabletypesize{\scriptsize}
\tablewidth{0pt}
\tablecaption{List of model properties for all globular cluster models \label{table:params}}
\tablehead{
	\colhead{No.} &
    \colhead{$N$} &
    \colhead{$r_{\rm{G}}$} &
    \colhead{$w_o$} &
    \colhead{$f_b$} &
    \colhead{$Z$} &
    \colhead{$r_v$} &
    \multicolumn{2}{c}{BH-formation kick} &    
    \multicolumn{3}{c}{High Mass Binaries} &
    \colhead{\rm{IMF} range} &
    \colhead{Winds} &
    \colhead{$N_{\rm{BH-MTB}}$} &
    \colhead{$N_{\rm{NS-MTB}}$} \\
    \cline{8-12}\\
    \colhead{} &
    \colhead{$(10^5$)} &
    \colhead{$(\rm{kpc})$} &
    \colhead{} &
    \colhead{} &
    \colhead{} &
    \colhead{$(\rm{pc})$} &
    \colhead{$\frac{\sigma_{\rm{BH}}}{\sigma_{\rm{NS}}}$} &
    \colhead{\texttt{FB}} &
    \colhead{$f_{b,\rm{high}}$} &
    \colhead{$q$ range} &
    \colhead{$\frac{dn}{d\, \log P}$} &
    \colhead{} &
    \colhead{} &
    \colhead{} &
    \colhead{}
}
\startdata
1 &  10 & 8 & 5 & 0.1 & 0.001 & 1 & 1.0 & n & 0.1 & $0.1/m_p,\,1$ & $P^0$ & $[0.1,100]$ & W & 0 & 13 \\
%$2^{*}$ &  10 & 8 & 5 & 0.1 & 0.001 & 1 & 1.0 & y & 0.1 & $0.1/m_p,\,1$ & $P^0$ & 1 & 0 \\
%$3^{*}$ &  10 & 8 & 5 & 0.1 & 0.001 & 1 & 1.0 & y & 0.1 & $0.1/m_p,\,1$ & $P^0$ & 1 & 1 \\
2 & 2 & 20 & 5 & 0.1 & 0.0002 & 1 & 1.0 & y & 0.1 & $0.1/m_p,\,1$ & $P^0$ & $[0.1,100]$ & W &0 & 1 \\
3 & 2 & 20 & 5 & 0.1 & 0.0002 & 1 & 1.0 & y & 0.1 & $0.1/m_p,\,1$ & $P^0$ & $[0.1,100]$ & W &0 & 1 \\
4 & 5 & 20 & 5 & 0.1 & 0.0002 & 1 & 1.0 & y & 0.1 & $0.1/m_p,\,1$ & $P^0$ & $[0.1,100]$ & W & 0 & 2 \\
5 & 5 & 20 & 5 & 0.1 & 0.0002 & 1 & 1.0 & y & 0.1 & $0.1/m_p,\,1$ & $P^0$ & $[0.1,100]$ & W & 2 & 1 \\
6 & 10 & 20 & 5 & 0.1 & 0.0002 & 1 & 1.0 & y & 0.1 & $0.1/m_p,\,1$ & $P^0$ & $[0.1,100]$ & W & 3 & 3 \\
7 & 10 & 20 & 5 & 0.1 & 0.0002 & 1 & 1.0 & y & 0.1 & $0.1/m_p,\,1$ & $P^0$ & $[0.1,100]$ & W & 2 & 2 \\
8 & 20 & 20 & 5 & 0.1 & 0.0002 & 1 & 1.0 & y & 0.1 & $0.1/m_p,\,1$ & $P^0$ & $[0.1,100]$ & W & 2 & 4 \\
9 & 20 & 20 & 5 & 0.1 & 0.0002 & 1 & 1.0 & y & 0.1 & $0.1/m_p,\,1$ & $P^0$ & $[0.1,100]$ & W & 4 & 3 \\
10 & 2 & 2 & 5 & 0.1 & 0.005 & 1 & 1.0 & y & 0.1 & $0.1/m_p,\,1$ & $P^0$ & $[0.1,100]$ & W & 1 & 1 \\
11 & 2 & 2 & 5 & 0.1 & 0.005 & 1 & 1.0 & y & 0.1 & $0.1/m_p,\,1$ & $P^0$ & $[0.1,100]$ & W & 0 & 2 \\
12 & 5 & 2 & 5 & 0.1 & 0.005 & 1 & 1.0 & y & 0.1 & $0.1/m_p,\,1$ & $P^0$ & $[0.1,100]$ & W & 0 & 2 \\
13 & 5 & 2 & 5 & 0.1 & 0.005 & 1 & 1.0 & y & 0.1 & $0.1/m_p,\,1$ & $P^0$ & $[0.1,100]$ & W & 0 & 2 \\
14 & 10 & 2 & 5 & 0.1 & 0.005 & 1 & 1.0 & y & 0.1 & $0.1/m_p,\,1$ & $P^0$ & $[0.1,100]$ & W & 0 & 2 \\
15 & 10 & 2 & 5 & 0.1 & 0.005 & 1 & 1.0 & y & 0.1 & $0.1/m_p,\,1$ & $P^0$ & $[0.1,100]$ & W & 0 & 2 \\
16 & 20 & 2 & 5 & 0.1 & 0.005 & 1 & 1.0 & y & 0.1 & $0.1/m_p,\,1$ & $P^0$ & $[0.1,100]$ & W & 0 & 2 \\
17 & 20 & 2 & 5 & 0.1 & 0.005 & 1 & 1.0 & y & 0.1 & $0.1/m_p,\,1$ & $P^0$ & $[0.1,100]$ & W & 3 & 5 \\
18 & 2 & 8 & 5 & 0.1 & 0.001 & 1 & 1.0 & y & 0.1 & $0.1/m_p,\,1$ & $P^0$ & $[0.1,100]$ & W & 0 & 0 \\
19 & 2 & 8 & 5 & 0.1 & 0.001 & 1 & 1.0 & y & 0.1 & $0.1/m_p,\,1$ & $P^0$ & $[0.1,100]$ & W & 0 & 0 \\
20 & 5 & 8 & 5 & 0.1 & 0.001 & 1 & 1.0 & y & 0.1 & $0.1/m_p,\,1$ & $P^0$ & $[0.1,100]$ & W & 1 & 1 \\
21 & 5 & 8 & 5 & 0.1 & 0.001 & 1 & 1.0 & y & 0.1 & $0.1/m_p,\,1$ & $P^0$ & $[0.1,100]$ & W & 1 & 1 \\
22 & 10 & 8 & 5 & 0.1 & 0.001 & 1 & 1.0 & y & 0.1 & $0.1/m_p,\,1$ & $P^0$ & $[0.1,100]$ & W & 0 & 2 \\
23 & 10 & 8 & 5 & 0.1 & 0.001 & 1 & 1.0 & y & 0.1 & $0.1/m_p,\,1$ & $P^0$ & $[0.1,100]$ & W & 1 & 2 \\
24 & 20 & 8 & 5 & 0.1 & 0.001 & 1 & 1.0 & y & 0.1 & $0.1/m_p,\,1$ & $P^0$ & $[0.1,100]$ & W & 1 & 2 \\
25 & 20 & 8 & 5 & 0.1 & 0.001 & 1 & 1.0 & y & 0.1 & $0.1/m_p,\,1$ & $P^0$ & $[0.1,100]$ & W & 5 & 2 \\
26 & 2 & 20 & 5 & 0.1 & 0.0002 & 2 & 1.0 & y & 0.1 & $0.1/m_p,\,1$ & $P^0$ & $[0.1,100]$ & W & 0 & 0 \\
27 & 2 & 20 & 5 & 0.1 & 0.0002 & 2 & 1.0 & y & 0.1 & $0.1/m_p,\,1$ & $P^0$ & $[0.1,100]$ & W & 0 & 1 \\
28 & 5 & 20 & 5 & 0.1 & 0.0002 & 2 & 1.0 & y & 0.1 & $0.1/m_p,\,1$ & $P^0$ & $[0.1,100]$ & W & 1 & 1 \\
29 & 5 & 20 & 5 & 0.1 & 0.0002 & 2 & 1.0 & y & 0.1 & $0.1/m_p,\,1$ & $P^0$ & $[0.1,100]$ & W & 0 & 1 \\
30 & 10 & 20 & 5 & 0.1 & 0.0002 & 2 & 1.0 & y & 0.1 & $0.1/m_p,\,1$ & $P^0$ & $[0.1,100]$ & W & 1 & 0 \\
31 & 10 & 20 & 5 & 0.1 & 0.0002 & 2 & 1.0 & y & 0.1 & $0.1/m_p,\,1$ & $P^0$ & $[0.1,100]$ & W & 1 & 1 \\
32 & 20 & 20 & 5 & 0.1 & 0.0002 & 2 & 1.0 & y & 0.1 & $0.1/m_p,\,1$ & $P^0$ & $[0.1,100]$ & W & 0 & 1 \\
33 & 20 & 20 & 5 & 0.1 & 0.0002 & 2 & 1.0 & y & 0.1 & $0.1/m_p,\,1$ & $P^0$ & $[0.1,100]$ & W & 0 & 1 \\
34 & 2 & 2 & 5 & 0.1 & 0.005 & 2 & 1.0 & y & 0.1 & $0.1/m_p,\,1$ & $P^0$ & $[0.1,100]$ & W & 0 & 0 \\
35 & 2 & 2 & 5 & 0.1 & 0.005 & 2 & 1.0 & y & 0.1 & $0.1/m_p,\,1$ & $P^0$ & $[0.1,100]$ & W & 0 & 0 \\
36 & 5 & 2 & 5 & 0.1 & 0.005 & 2 & 1.0 & y & 0.1 & $0.1/m_p,\,1$ & $P^0$ & $[0.1,100]$ & W & 2 & 1 \\
37 & 5 & 2 & 5 & 0.1 & 0.005 & 2 & 1.0 & y & 0.1 & $0.1/m_p,\,1$ & $P^0$ & $[0.1,100]$ & W & 0 & 2 \\
38 & 10 & 2 & 5 & 0.1 & 0.005 & 2 & 1.0 & y & 0.1 & $0.1/m_p,\,1$ & $P^0$ & $[0.1,100]$ & W & 0 & 2 \\
39 & 10 & 2 & 5 & 0.1 & 0.005 & 2 & 1.0 & y & 0.1 & $0.1/m_p,\,1$ & $P^0$ & $[0.1,100]$ & W & 1 & 3 \\
40 & 20 & 2 & 5 & 0.1 & 0.005 & 2 & 1.0 & y & 0.1 & $0.1/m_p,\,1$ & $P^0$ & $[0.1,100]$ & W & 0 & 8 \\
41 & 20 & 2 & 5 & 0.1 & 0.005 & 2 & 1.0 & y & 0.1 & $0.1/m_p,\,1$ & $P^0$ & $[0.1,100]$ & W & 0 & 5 \\
42 & 2 & 8 & 5 & 0.1 & 0.001 & 2 & 1.0 & y & 0.1 & $0.1/m_p,\,1$ & $P^0$ & $[0.1,100]$ & W & 0 & 0 \\
43 & 2 & 8 & 5 & 0.1 & 0.001 & 2 & 1.0 & y & 0.1 & $0.1/m_p,\,1$ & $P^0$ & $[0.1,100]$ & W & 0 & 0 \\
44 & 5 & 8 & 5 & 0.1 & 0.001 & 2 & 1.0 & y & 0.1 & $0.1/m_p,\,1$ & $P^0$ & $[0.1,100]$ & W & 0 & 1 \\
45 & 5 & 8 & 5 & 0.1 & 0.001 & 2 & 1.0 & y & 0.1 & $0.1/m_p,\,1$ & $P^0$ & $[0.1,100]$ & W & 0 & 2 \\
46 & 10 & 8 & 5 & 0.1 & 0.001 & 2 & 1.0 & y & 0.1 & $0.1/m_p,\,1$ & $P^0$ & $[0.1,100]$ & W & 0 & 1 \\
47 & 10 & 8 & 5 & 0.1 & 0.001 & 2 & 1.0 & y & 0.1 & $0.1/m_p,\,1$ & $P^0$ & $[0.1,100]$ & W & 1 & 0 \\
48 & 20 & 8 & 5 & 0.1 & 0.001 & 2 & 1.0 & y & 0.1 & $0.1/m_p,\,1$ & $P^0$ & $[0.1,100]$ & W & 1 & 6 \\
49 & 20 & 8 & 5 & 0.1 & 0.001 & 2 & 1.0 & y & 0.1 & $0.1/m_p,\,1$ & $P^0$ & $[0.1,100]$ & W & 0 & 2 \\
50 & 1 & 4.6 & 5 & 0.05 & 0.00055 & 1 & 1.0 & y & 0.05 & $0.1/m_p,\,1$ & $P^0$ & $[0.1,100]$ & S & 3 & 0 \\
51 & 2.4 & 4.6 & 5 & 0.05 & 0.00055 & 0.8 & 1.0 & y & 0.05 & $0.1/m_p,\,1$ & $P^0$ & $[0.1,100]$ & S & 5 & 0 \\
52 & 2.4 & 4.6 & 5 & 0.05 & 0.00055 & 1 & 1.0 & y & 0.05 & $0.1/m_p,\,1$ & $P^0$ & $[0.1,100]$ & S & 1 & 0 \\
53 & 2.6 & 4.6 & 3 & 0.05 & 0.00055 & 0.8 & 1.0 & y & 0.05 & $0.1/m_p,\,1$ & $P^0$ & $[0.1,100]$ & S & 2 & 0 \\
54 & 2.6 & 4.6 & 4.5 & 0.05 & 0.00055 & 0.8 & 1.0 & y & 0.05 & $0.1/m_p,\,1$ & $P^0$ & $[0.1,100]$ & S & 0 & 0 \\
55 & 2.6 & 4.6 & 4.8 & 0.05 & 0.00055 & 0.8 & 1.0 & y & 0.05 & $0.1/m_p,\,1$ & $P^0$ & $[0.1,100]$ & S & 3 & 0 \\
56 & 2.6 & 4.6 & 4 & 0.05 & 0.00055 & 0.8 & 1.0 & y & 0.05 & $0.1/m_p,\,1$ & $P^0$ & $[0.1,100]$ & S & 3 & 0 \\
57 & 2.6 & 4.6 & 5 & 0.05 & 0.00055 & 0.8 & 1.0 & y & 0.05 & $0.1/m_p,\,1$ & $P^0$ & $[0.1,100]$ & S & 7 & 1 \\
58 & 2.6 & 4.6 & 5 & 0.05 & 0.00055 & 1.2 & 1.0 & y & 0.05 & $0.1/m_p,\,1$ & $P^0$ & $[0.1,100]$ & S & 3 & 0 \\
59 & 2.6 & 4.6 & 5 & 0.05 & 0.00055 & 1 & 1.0 & y & 0.05 & $0.1/m_p,\,1$ & $P^0$ & $[0.1,100]$ & S & 2 & 0 \\
60 & 2 & 4.6 & 5 & 0.05 & 0.00055 & 1 & 1.0 & y & 0.05 & $0.1/m_p,\,1$ & $P^0$ & $[0.1,100]$ & S & 0 & 0 \\
61 & 3.25 & 4.6 & 5 & 0.05 & 0.00055 & 1 & 1.0 & y & 0.05 & $0.1/m_p,\,1$ & $P^0$ & $[0.1,100]$ & S & 5 & 1 \\
62 & 3.5 & 4.6 & 4 & 0.05 & 0.00055 & 1 & 1.0 & y & 0.05 & $0.1/m_p,\,1$ & $P^0$ & $[0.1,100]$ & S & 1 & 0 \\
63 & 3.5 & 4.6 & 5 & 0.05 & 0.00055 & 1 & 1.0 & y & 0.05 & $0.1/m_p,\,1$ & $P^0$ & $[0.1,100]$ & S & 2 & 0 \\
64 & 3.75 & 4.6 & 4 & 0.05 & 0.00055 & 1 & 1.0 & y & 0.05 & $0.1/m_p,\,1$ & $P^0$ & $[0.1,100]$ & S & 1 & 2 \\
65 & 3.75 & 4.6 & 5.5 & 0.05 & 0.00055 & 0.8 & 1.0 & y & 0.05 & $0.1/m_p,\,1$ & $P^0$ & $[0.1,100]$ & S & 6 & 0 \\
66 & 3.75 & 4.6 & 5 & 0.05 & 0.00055 & 0.8 & 1.0 & y & 0.05 & $0.1/m_p,\,1$ & $P^0$ & $[0.1,100]$ & S & 5 & 1 \\
67 & 3.75 & 4.6 & 5 & 0.05 & 0.00055 & 1 & 1.0 & y & 0.05 & $0.1/m_p,\,1$ & $P^0$ & $[0.1,100]$ & S & 1 & 2 \\
68 & 3.75 & 4.6 & 6 & 0.05 & 0.00055 & 1 & 1.0 & y & 0.05 & $0.1/m_p,\,1$ & $P^0$ & $[0.1,100]$ & S & 4 & 1 \\
69 & 3.8 & 4.6 & 5.1 & 0.05 & 0.00055 & 0.9 & 1.0 & y & 0.05 & $0.1/m_p,\,1$ & $P^0$ & $[0.1,100]$ & S & 3 & 1 \\
70 & 3.8 & 4.6 & 5.2 & 0.05 & 0.00055 & 0.85 & 1.0 & y & 0.05 & $0.1/m_p,\,1$ & $P^0$ & $[0.1,100]$ & S & 5 & 1 \\
71 & 3 & 4.6 & 4 & 0.05 & 0.00055 & 2 & 1.0 & y & 0.05 & $0.1/m_p,\,1$ & $P^0$ & $[0.1,100]$ & S & 1 & 0 \\
72 & 3 & 4.6 & 5 & 0.05 & 0.00055 & 1 & 1.0 & y & 0.05 & $0.1/m_p,\,1$ & $P^0$ & $[0.1,100]$ & S & 1 & 0 \\
73 & 4 & 4.6 & 5 & 0.05 & 0.00055 & 0.8 & 1.0 & y & 0.05 & $0.1/m_p,\,1$ & $P^0$ & $[0.1,100]$ & S & 3 & 0 \\
74 & 4 & 4.6 & 5 & 0.05 & 0.00055 & 1 & 1.0 & y & 0.05 & $0.1/m_p,\,1$ & $P^0$ & $[0.1,100]$ & S & 1 & 0 \\
75 & 4 & 4.6 & 6 & 0.05 & 0.00055 & 1 & 1.0 & y & 0.05 & $0.1/m_p,\,1$ & $P^0$ & $[0.1,100]$ & S & 4 & 1 \\
76 & 6 & 4.6 & 4 & 0.05 & 0.00055 & 1 & 1.0 & y & 0.05 & $0.1/m_p,\,1$ & $P^0$ & $[0.1,100]$ & S & 1 & 0 \\
77 & 6 & 4.6 & 5 & 0.05 & 0.00055 & 1 & 1.0 & y & 0.05 & $0.1/m_p,\,1$ & $P^0$ & $[0.1,100]$ & S & 0 & 0 \\
78 & 6 & 4.6 & 6 & 0.05 & 0.00055 & 1 & 1.0 & y & 0.05 & $0.1/m_p,\,1$ & $P^0$ & $[0.1,100]$ & S & 1 & 2 \\
79 & 8 & 4.6 & 5 & 0.05 & 0.00055 & 1 & 1.0 & y & 0.05 & $0.1/m_p,\,1$ & $P^0$ & $[0.1,100]$ & S & 1 & 0 \\
80 & 8 & 8 & 5 & 0.05 & 0.001 & 2 & 1 & y & 0.05 & $0.1/m_p,\,1$ & $P^0$ & $[0.1,100]$ & S & 0 & 3 \\
81 & 8 & 8 & 5 & 0.05 & 0.001 & 2 & 1 & n & 0.05 & $0.1/m_p,\,1$ & $P^0$ & $[0.1,100]$ & S & 0 & 10 \\
82 & 8 & 8 & 5 & 0.05 & 0.001 & 2 & 0.1 & n & 0.05 & $0.1/m_p,\,1$ & $P^0$ & $[0.1,100]$ & S & 0 & 3 \\
83 & 8 & 8 & 5 & 0.05 & 0.001 & 2 & 0.01 & n & 0.05 & $0.1/m_p,\,1$ & $P^0$ & $[0.1,100]$ & S & 0 & 2 \\
84 & 8 & 8 & 5 & 0.05 & 0.001 & 2 & 1 & y & 0.05 & $0.1/m_p,\,1$ & $P^0$ & $[0.1,100]$ & S & 0 & 0 \\
85 & 8 & 8 & 5 & 0.05 & 0.001 & 2 & 1 & n & 0.05 & $0.1/m_p,\,1$ & $P^0$ & $[0.1,100]$ & S & 0 & 3 \\
86 & 8 & 8 & 5 & 0.05 & 0.001 & 2 & 0.1 & n & 0.05 & $0.1/m_p,\,1$ & $P^0$ & $[0.1,100]$ & S & 0 & 2 \\
87 & 8 & 8 & 5 & 0.05 & 0.001 & 2 & 0.01 & n & 0.05 & $0.1/m_p,\,1$ & $P^0$ & $[0.1,100]$ & S & 0 & 1 \\
88 & 8 & 8 & 5 & 0.05 & 0.001 & 2 & 1 & y & 0.05 & $0.1/m_p,\,1$ & $P^0$ & $[0.1,100]$ & S & 0 & 0 \\
89 & 8 & 8 & 5 & 0.05 & 0.001 & 2 & 0.1 & n & 0.05 & $0.1/m_p,\,1$ & $P^0$ & $[0.1,100]$ & S & 0 & 4 \\
90 & 8 & 8 & 5 & 0.05 & 0.001 & 2 & 0.01 & n & 0.05 & $0.1/m_p,\,1$ & $P^0$ & $[0.1,100]$ & S & 1 & 1 \\
91 & 8 & 8 & 5 & 0.05 & 0.001 & 2 & 1 & y & 0.7 & $[0.6,1]$ & $P^{-0.55}$  & $[0.1,100]$ & S & 0 & 1 \\
92 & 8 & 8 & 5 & 0.05 & 0.001 & 2 & 1 & n & 0.7 & $[0.6,1]$ & $P^{-0.55}$  & $[0.1,100]$ & S & 0 & 1 \\
93 & 8 & 8 & 5 & 0.05 & 0.001 & 2 & 0.1 & n & 0.7 & $[0.6,1]$ & $P^{-0.55}$  & $[0.1,100]$ & S & 0 & 2 \\
94 & 8 & 8 & 5 & 0.05 & 0.001 & 2 & 0.01 & n & 0.7 & $[0.6,1]$ & $P^{-0.55}$  & $[0.1,100]$ & S & 0 & 0 \\
95 & 8 & 8 & 5 & 0.05 & 0.001 & 2 & 1 & y & 0.05 &  $0.1/m_p,\,1$ & $P^0$ & $[0.1,100]$ & S & 2 & 2 \\
96 & 8 & 8 & 5 & 0.05 & 0.001 & 2 & 1 & n & 0.05 &  $0.1/m_p,\,1$ & $P^0$ & $[0.1,100]$ & S & 0 & 3 \\
97 & 8 & 8 & 5 & 0.05 & 0.001 & 2 & 0.01 & n & 0.05 &  $0.1/m_p,\,1$ & $P^0$ & $[0.1,100]$ & S & 0 & 1 \\
98 & 8 & 8 & 5 & 0.05 & 0.001 & 2 & 0.01 & n & 1 & $[0.1/m_p,\,1]$ & $P^0$ & $[0.1,100]$ & S & 0 & 1 \\
99 & 8 & 8 & 5 & 0.05 & 0.001 & 2 & 0.01 & n & 0 & $-$ & $-$ & $[0.1,100]$ & S & 2 & 2 \\
100 & 8 & 8 & 5 & 0.05 & 0.001 & 2 & 1 & y & 0.7 & $[0.6,1]$ & $P^{-0.55}$  & $[0.1,100]$ & S & 0 & 1 \\
101 & 8 & 8 & 5 & 0.05 & 0.001 & 2 & 1 & n & 0.7 & $[0.6,1]$ & $P^{-0.55}$  & $[0.1,100]$ & S & 0 & 2 \\
102 & 8 & 8 & 5 & 0.05 & 0.001 & 2 & 0.1 & n & 0.7 & $[0.6,1]$ & $P^{-0.55}$  & $[0.1,100]$ & S & 0 & 2 \\
103 & 8 & 8 & 5 & 0.05 & 0.001 & 2 & 0.01 & n & 0.7 & $[0.6,1]$ & $P^{-0.55}$  & $[0.1,100]$ & S & 0 & 1 \\
104 & 8 & 8 & 5 & 0.05 & 0.001 & 2 & 1 & y & 0.7 & $[0.6,1]$ & $P^{-0.55}$  & $[0.1,100]$ & S & 0 & 1 \\
105 & 8 & 8 & 5 & 0.05 & 0.001 & 2 & 1 & n & 0.7 & $[0.6,1]$ & $P^{-0.55}$  & $[0.1,100]$ & S & 0 & 1 \\
106 & 8 & 8 & 5 & 0.05 & 0.001 & 2 & 0.01 & n & 0.7 & $[0.6,1]$ & $P^{-0.55}$  & $[0.1,100]$ & S & 0 & 0 \\
107 & 8 & 4 & 5 & 0.05 & 0.001 & 2 & 1 & y & 0.7 & $[0.6,1]$ & $P^{-0.55}$  & $[0.1,100]$ & S & 0 & 2 \\
108 & 8 & 4 & 5 & 0.05 & 0.001 & 2 & 1 & n & 0.7 & $[0.6,1]$ & $P^{-0.55}$  & $[0.1,100]$ & S & 0 & 1 \\
109 & 8 & 8 & 5 & 0.05 & 0.001 & 2 & 1 & y & 0.05 & $[0.1/m_p,\,1]$ & $P^0$ & $[0.08,150]$ & W & 3 & 2 \\
110 & 8 & 8 & 5 & 0.05 & 0.001 & 2 & 1 & y & 0.05 & $[0.1/m_p,\,1]$ & $P^0$ & $[0.08,150]$ & W & 1 & 1 \\
111 & 8 & 8 & 5 & 0.05 & 0.001 & 2 & 1 & y & 0.05 & $[0.1/m_p,\,1]$ & $P^0$ & $[0.08,150]$ & W & 1 & 2 \\
112 & 8 & 8 & 5 & 0.05 & 0.001 & 2 & 1 & y & 0.05 & $[0.1/m_p,\,1]$ & $P^0$ & $[0.08,150]$ & W & 0 & 4 \\
113 & 8 & 8 & 5 & 0.05 & 0.001 & 1 & 1 & y & 0.05 & $[0.1/m_p,\,1]$ & $P^0$ & $[0.08,150]$ & W & 4 & 2 \\
114 & 2 & 8 & 5 & 0.05 & 0.001 & 1 & 1 & y & 0.05 & $[0.1/m_p,\,1]$ & $P^0$ & $[0.08,150]$ & W & 6 & 4 \\
115 & 2 & 8 & 5 & 0.05 & 0.001 & 1 & 1 & y & 0.05 & $[0.1/m_p,\,1]$ & $P^0$ & $[0.08,150]$ & W & 5 & 2 \\
116 & 2 & 8 & 5 & 0.05 & 0.001 & 2 & 1 & y & 0.05 & $[0.1/m_p,\,1]$ & $P^0$ & $[0.08,150]$ & W & 1 & 2 \\
117 & 2 & 8 & 5 & 0.05 & 0.001 & 2 & 1 & y & 0.05 & $[0.1/m_p,\,1]$ & $P^0$ & $[0.08,150]$ & W & 2 & 0 \\
118 & 2 & 8 & 5 & 0.05 & 0.001 & 2 & 1 & y & 0.05 & $[0.1/m_p,\,1]$ & $P^0$ & $[0.08,150]$ & W & 2 & 0 \\
119 & 2 & 8 & 5 & 0.05 & 0.001 & 2 & 1 & y & 0.05 & $[0.1/m_p,\,1]$ & $P^0$ & $[0.08,150]$ & W & 5 & 2 \\
120 & 2 & 8 & 5 & 0.05 & 0.001 & 2 & 1 & y & 0.05 & $[0.1/m_p,\,1]$ & $P^0$ & $[0.08,150]$ & W & 0 & 3 \\
121 & 2 & 8 & 5 & 0.05 & 0.001 & 2 & 1 & y & 0.05 & $[0.1/m_p,\,1]$ & $P^0$ & $[0.08,150]$ & W & 2 & 0 \\
122 & 8 & 8 & 5 & 0.05 & 0.001 & 1 & 1 & y & 0.05 & $[0.1/m_p,\,1]$ & $P^0$ & $[0.08,150]$ & W & 5 & 1 \\
123 & 8 & 8 & 5 & 0.05 & 0.001 & 2 & 1 & y & 0.05 & $[0.1/m_p,\,1]$ & $P^0$ & $[0.1,100]$ & W & 1 & 1 \\
124 & 8 & 8 & 5 & 0.05 & 0.005 & 2 & 1 & y & 0.05 & $[0.1/m_p,\,1]$ & $P^0$ & $[0.08,150]$ & W & 1 & 3 \\
125 & 8 & 8 & 5 & 0.05 & 0.005 & 2 & 1 & y & 0.05 & $[0.1/m_p,\,1]$ & $P^0$ & $[0.08,150]$ & W & 0 & 2 \\
126 & 8 & 8 & 5 & 0.05 & 0.005 & 2 & 1 & y & 0.05 & $[0.1/m_p,\,1]$ & $P^0$ & $[0.08,150]$ & W & 2 & 0 \\
127 & 8 & 8 & 5 & 0.05 & 0.005 & 2 & 1 & y & 0.05 & $[0.1/m_p,\,1]$ & $P^0$ & $[0.08,150]$ & W & 1 & 4 \\
128 & 8 & 8 & 5 & 0.05 & 0.005 & 2 & 1 & y & 0.05 & $[0.1/m_p,\,1]$ & $P^0$ & $[0.08,150]$ & W & 0 & 2 \\
129 & 8 & 8 & 5 & 0.05 & 0.0001 & 2 & 1 & y & 0.05 & $[0.1/m_p,\,1]$ & $P^0$ & $[0.08,150]$ & W & 2 & 18 \\
130 & 8 & 8 & 5 & 0.05 & 0.0001 & 2 & 1 & y & 0.05 & $[0.1/m_p,\,1]$ & $P^0$ & $[0.08,150]$ & W & 3 & 24 \\
131 & 8 & 8 & 5 & 0.05 & 0.0001 & 2 & 1 & y & 0.05 &$[0.1/m_p,\,1]$ & $P^0$ & $[0.08,150]$ & W & 0 & 20 \\
132 & 8 & 8 & 5 & 0.05 & 0.0001 & 2 & 1 & y & 0.05 & $[0.1/m_p,\,1]$ & $P^0$ & $[0.08,150]$ & W & 0 & 17 \\
133 & 8 & 8 & 5 & 0.05 & 0.0005 & 2 & 1 & y & 0.05 & $[0.1/m_p,\,1]$ & $P^0$ & $[0.08,150]$ & W & 2 & 2 \\
134 & 8 & 8 & 5 & 0.05 & 0.0005 & 2 & 1 & y & 0.05 & $[0.1/m_p,\,1]$ & $P^0$ & $[0.08,150]$ & W & 2 & 1 \\
135 & 8 & 8 & 5 & 0.05 & 0.0005 & 2 & 1 & y & 0.05 &$[0.1/m_p,\,1]$ & $P^0$ & $[0.08,150]$ & W & 6 & 1 \\
136 & 8 & 8 & 5 & 0.05 & 0.0005 & 2 & 1 & y & 0.05 & $[0.1/m_p,\,1]$ & $P^0$ & $[0.08,150]$ & W & 2 & 1 \\
137 & 8 & 8 & 5 & 0.05 & 0.0005 & 2 & 1 & y & 0.05 & $[0.1/m_p,\,1]$ & $P^0$ & $[0.08,150]$ & W & 0 & 1
\enddata
\tablecomments{Relevant properties of all GC models used in this study. Each initial parameter is described in Section \ref{sec:model_properties}. The final two columns show the total number of mass-transferring BH binaries ($N_{\rm{BH-MTB}}$) and mass-transferring NS binaries ($N_{\rm{NS-MTB}}$) ejected from each model by 12 Gyr.}
\end{deluxetable*}

.

\end{document}